# Review on Modeling of Mechanical and Thermal Properties of Nano- and Micro-Composites

S. I. Kundalwal

*Discipline of Mechanical Engineering, Indian Institute of Technology, Indore, 453552, India*

**Abstract**

This article deals with the prediction of thermomechanical properties of fiber reinforced composites using several micromechanics models. These include strength of material approach, Halpin-Tsai equations, multi-phase mechanics of materials approaches, multi-phase Mori-Tanaka models, composite cylindrical assemblage model, Voigt-Reuss models, modified mixture rule, Cox model, effective medium approach and method of cells. Several composite systems reinforced with short and long, aligned, random and wavy reinforcements were considered. In addition, different aspects such as fiber-matrix interphase, fiber-matrix interfacial thermal resistance, fiber geometry, and multiple types of reinforcements were considered to model the composites systems. The current study also presents some important preliminary concepts and application of developed micromechanics models to advanced nanocomposites such as carbon nanotube reinforced composite. Main contribution of the current work is the investigation of several analytical micromechanical models, while most of the existing studies on the subject deal with only one or two approaches considering few aspects.

Corresponding author. Tel.: +91-8830670644
E-mail address: kundalwal@iiti.ac.in (S.I. Kundalwal).

# Contents

# LIST OF ABBREVIATIONS AND SYMBOLS

| | |
|---|---|
| CCA | Cylinder assemblage approach |
| CTE | Coefficient of thermal expansion |
| MOC | Method of cells |
| MOM | Mechanics of materials |
| MT | Mori-Tanaka |
| RVE | Representative volume element |
| SMC | Sheet moulding compound |
| A | Amplitude of the fiber (m) |
| $a_k$ | Kaptiza radius (m) |
| [A] | Matrix formed by the geometrical parameters and the material properties of the subcell |
| $[A_r]$ | Strain concentration tensor for the $r^{th}$ phase |
| $[A_c]$ | Matrix containing strain concentration factors of all the subcells |
| $\left[A_c^{\beta\gamma}\right]$ | Strain concentration matrix of the $\beta\gamma^{th}$ subcell |
| $[A_G]$ | Matrix constructed by the geometrical parameters of all the subcells |
| $[A_m]$ | Matrix containing the elastic properties of all the subcells |
| b | Width of the subcell |
| [B] | Matrix constructed by the geometrical parameters of the cell |
| $[\bar{C}^C]$ | Elastic coefficient matrix of the composite containing wavy fibers (GPa) |
| $[C^r]$ | Elastic coefficient matrix of the $r^{th}$ phase (GPa) |
| $[C^{\beta\gamma}]$ | Elastic coefficient matrix of the $\beta\gamma^{th}$ subcell (GPa) |
| $C_{ij}^r$ | Elastic coefficients of the $r^{th}$ phase (GPa) |
| $C_{ij}^{\beta\gamma}$ | Elastic coefficients of the $\beta\gamma^{th}$ subcell (GPa) |

| $d_f$ | Diameter of the fiber (m) |
|---|---|
| $[D_c]$ | Thermal concentration matrix of all the subcells |
| $[D_c^{\beta\gamma}]$ | Thermal concentration matrix of the $\beta\gamma^{th}$ subcell |
| $E_A$ and $E_T$ | Effective axial and transverse Young's moduli of the composite containing aligned fibers, respectively (GPa) |
| $E^c$ and $\nu^c$ | Young's modulus and Poisson's ratio of a two-dimensional randomly oriented composite |
| $E^r$ | Young's modulus of the $r^{th}$ phase (GPa) |
| $E_1, E_2, E_3$ | Young's modulus referred to the principal material coordinate (1–2–3) axis (GPa) |
| $G^r$ | Shear modulus of the $r^{th}$ phase (GPa) |
| $G_{12}, G_{13}$ | Axial shear moduli (GPa) |
| $G_{23}$ | Transverse shear modulus (GPa) |
| $h$ | Height of the subcell (m) |
| $[I]$ | Fourth order identity matrix |
| $K_{23}$ | Effective bulk modulus (GPa) |
| $K^r$ | Thermal conductivity of the rth phase (W/mK) |
| $l$ | Length of the unit cell (m) |
| $L_f$ | Length of the fiber |
| $L_{nr}$ | Running length of the sinusoidally wavy fiber (m) |
| $L_{RVE}$ | Length of the RVE (m) |
| $M$ | Number of subcells present in the cell along the 1– direction |
| $[M]$ | Strain concentration tensor |
| $[\hat{M}]$ | Alternate strain tensor that relate the average fiber strain to the average matrix strain |
| $N$ | Number of subcells present in the cell along the 2– direction |
| [N] | Stress concentration tensor |
| n | Number of waves of the sinusoidally wavy fiber along its axial direction |
| {n} | Unit vector |
| $K_1, K_2, K_3$ | Thermal conductivity referred to the principal material coordinate (1–2–3) axis (W/mK) |
| $R_k$ | Interfacial thermal resistance between the fiber and matrix (m²K/W) |
| $r_r$ | Radius of the $r^{th}$ phase (m) |
| $[S^r]$ | Elastic compliance matrix of the $r^{th}$ phase |
| $S_{ij}^r$ | Elements of compliance matrix of the $r^{th}$ phase |
| [S] | Eshelby tensor |
| $S_{ijkl}$ | Elements of Eshelby tensor |
| [T] | Transformation matrix |
| V | Volume of the RVE of the composite |
| $V^r$ | Volume of the $r^{th}$ phase |
| $v_r$ | Volume fraction of the of the $r^{th}$ phase |
| $V_{\beta\gamma}$ | Volume of the $\beta\gamma^{th}$ subcell (m³) |
| $w_r$ | Weight fraction of the $r^{th}$ phase (kg) |
| $\{\alpha\}$ | Thermal expansion coefficient vector of the composite (K⁻¹) |
| $\{\alpha^r\}$ | Thermal expansion coefficient vector of the $r^{th}$ phase (K⁻¹) |
| $\alpha_i$ | Thermal expansion coefficients of the composite (K⁻¹) |
| $\{\alpha^{\beta\gamma}\}$ | Thermal expansion coefficient vector of the $\beta\gamma^{th}$ subcell (K⁻¹) |
| $\Delta T$ | Temperature deviation from the reference temperature (K) |
| $\{\varepsilon\}$ | Strain vector |
| $\{\bar{\varepsilon}\}$ | Volume average strain vector |
| $\{\varepsilon^*\}$ | Eigenstrain vector |
| $\{\varepsilon^0\}$ | Uniform strain vector |
| $\{\varepsilon^{per}\}$ | Perturbation strain vector |
| $\{\varepsilon^r\}$ | Strain vector of the $r^{th}$ phase |
| $\{\varepsilon_T\}$ | Total strain vector |
| $\{\epsilon^{\beta\gamma}\}$ | Strain vector of the $\beta\gamma^{th}$ subcell |
| $\varepsilon_1^r, \varepsilon_2^r, \varepsilon_3^r$ | Normal strains along the principal material coordinate axes 1, 2 and 3, respectively, in the $r^{th}$ phase |
| $\varepsilon_{12}^r, \varepsilon_{13}^r, \varepsilon_{23}^r$ | Shearing strains in the $r^{th}$ phase |
| $\eta$ | Adhesion exponent |
| $\lambda$ | Wavelength of the fiber (m) |
| $\nu_{12}, \nu_{13}, \nu_{23}$ | Poisson's ratios |
| $\nu^f, \nu^i \ \& \ \nu^p$ | Poisson's ratios of the fiber, interphase and matrix, respectively |
| $\xi$ | Reinforcing factor |
| $\rho_r$ | Density of the $r^{th}$ phase (kg/ m$^3$) |
| $\{\sigma\}$ | Stress vector (GPa) |
| $\{\bar{\sigma}\}$ | Volume average stress vector (GPa) |
| $\{\sigma^0\}$ | Uniform stress vector (GPa) |
| $\{\sigma^r\}$ | Stress vector of the $r^{th}$ phase (GPa) |
| $\{\sigma^{\beta\gamma}\}$ | Stress vector of the $\beta\gamma^{th}$ subcell (GPa) |
| $\sigma_1^r, \sigma_2^r, \sigma_3^r$ | Normal stresses along the principal material coordinate axes 1, 2 and 3, respectively, in the $r^{th}$ phase (GPa) |
| $\sigma_{12}^r, \sigma_{13}^r, \sigma_{23}^r$ | Shearing stresses in the $r^{th}$ phase (GPa) |
| $\omega$ | Wave frequency of the fiber (m$^{-1}$) |

# INTRODUCTION

A composite is a structural material that consists of two or more constituents, combined at a macroscopic level and are not soluble in each other. One constituent is called the reinforcing phase and the one in which it is embedded is called the matrix. The reinforcing phase material may be in the form of fibers, particles, or flakes. Examples of composite systems include concrete reinforced with steel and epoxy reinforced with fibers. In addition, advanced composites are being extensively used in the aerospace industries. These composites have high performance reinforcements of a thin diameter in a matrix material such as epoxy-aluminium, graphite-epoxy, Kevlar-epoxy and boron-aluminium composites. These materials have now found applications in commercial industries as well [1]. There are many advantages of using composite materials over traditional materials such as the possibility of weight or volume reduction in a structure while maintaining a comparable or improved performance level.

Over the last five decades, substantial progress has been made in the micromechanics of composite materials. One major objective of micromechanics of heterogeneous materials is to determine the effective overall properties in terms of the properties of the constituents and their microstructure. The most common properties are the thermoelastic coefficients and thermal conductivities. As outlined by McCullough [2, 3], numerous micromechanical models have been developed to predict the macroscopic behavior of polymeric composite materials reinforced with typical reinforcements such as carbon or glass fibers. These micromechanical models assume that the fiber, matrix, and sometimes, the interface, are continuous materials and the constitutive equations for the bulk composite material are formulated based on assumptions of continuum mechanics.

Several micromechanical models have been developed by researchers to predict the effective properties of composites. The earliest works of Voigt [4] and Reuss [5] were primarily concerned with polycrystals. Their models can be applied to fiber reinforced composites. Voigt assumed that the states of strains are constant in the constituents under the application of load. On the other hand, Reuss assumed that the states of stresses are constant in the constituents. Using Voigt and Reuss theories, Hill [6] derived upper and lower bounds on the effective properties of a composite material. Likewise, Hashin and Shtrikman [7] derived upper and lower bounds for the effective elastic properties of quasi-isotropic and quasi-homogeneous multiphase materials using a variational approach. Hashin and Rosen [8] derived bounds and expressions for the effective elastic moduli of composite materials reinforced with aligned hollow circular fibers using a variational method. They obtained the exact results for hexagonal arrays of identical fibers and approximate results for random array of fibers. Hill [9, 10] determined the macroscopic elastic moduli of two-phase composites and showed that the main overall elastic moduli of composites with transversely isotropic phases are connected by simple universal relations which are independent of the geometry at given volume fraction. Chamis and Sendeckyj [11] presented a critique on the theories predicting the thermoelastic properties of unidirectional fiber reinforced composites. The theories were considered by them include netting analysis, mechanics of materials, self-consistent model, variational, exact, statistical, discrete element and semi empirical methods. Several researchers used the micromechanics-based Mori–Tanaka method [12] to predict the effective elastic properties of a composite material. This method has been successfully applied to transversely isotropic inclusions by Qui and Weng [13]. Aboudi [14] derived method of cells approaches to predict the thermoelastic and viscoelastic properties of composites. Several other micromechanical models have also been developed to determine the overall properties of composites; these include the self-consistent scheme [15, 16], the generalized self-consistent scheme [17], the differential schemes [18, 19], a stepping scheme for multi-inclusion composites [20], two and three-phase mechanics of materials approaches [21-23], micromechanical damage models [24], application of multi-phase Mori-Tanaka models [25-29], and pull-out and shear lag models [30-32].

The method of micromechanical approaches varies from simple analysis to complex ones. Limited studies have been reported on the use of most established micromechanical techniques. These studies were either restricted to few micromechanics models or reported selected properties taking in to account only special cases. Therefore, in this study, endeavour has been made to present variety of micromechanics techniques characterizing the effective thermoelastic and thermal properties of composites considering different influencing parameters such as fiber structure and orientation, fiber waviness, fiber-matrix interphase, fiber-matrix interfacial thermal resistance, and multiple types of reinforcements. This study provides a concise description and evaluation of established micromechanical models which are instructive in the prediction of thermoelastic properties and thermal properties. The current study provides important insight in better understanding and judicious application of these techniques as well as quick and intuitive predictions for fiber reinforced composites. Preliminary concepts, typical results and current trends in nanocomposites are also presented and discussed herein.

**BASIC EQUATIONS**

Consider a lamina of fiber reinforced composite material as shown in Fig. 1. The 1-2-3 orthogonal coordinate system is used where the directions are taken as follows:

- The 1-axis is aligned with the fiber axis,
- The 2-axis is in the plane of a lamina and perpendicular to the fibers, and
- The 3-axis is perpendicular to the plane of a lamina and thus also perpendicular to the fibers.

The 1-axis is also called the fiber axis, while the 2- and 3-axes are called the matrix directions. This 1-2-3 coordinate system is called the principal material coordinate system. The states of stresses and strains in the lamina will be referred to the principal material coordinate system.

At this level of analysis, the states of stress and strain of an individual constituent are not considered. The effect of the fiber reinforcement is smeared over the volume of the lamina. It is assumed that the fiber-matrix system is replaced by a single homogenous material. Obviously, this single material with different properties in three mutually perpendicular directions is called as orthotropic material. The Hook's law for such materials is given by

$$\begin{Bmatrix} \varepsilon_{11} \\ \varepsilon_{22} \\ \varepsilon_{33} \\ \varepsilon_{23} \\ \varepsilon_{13} \\ \varepsilon_{12} \end{Bmatrix} = \begin{bmatrix} 1/E_1 & -v_{21}/E_1 & -v_{31}/E_3 & 0 & 0 & 0 \\ -v_{12}/E_1 & 1/E_2 & -v_{32}/E_3 & 0 & 0 & 0 \\ -v_{13}/E_1 & -v_{23}/E_2 & 1/E_3 & 0 & 0 & 0 \\ 0 & 0 & 0 & 1/G_{32} & 0 & 0 \\ 0 & 0 & 0 & 0 & 1/G_{13} & 0 \\ 0 & 0 & 0 & 0 & 0 & 1/G_{12} \end{bmatrix} \begin{Bmatrix} \sigma_{11} \\ \sigma_{22} \\ \sigma_{33} \\ \sigma_{23} \\ \sigma_{13} \\ \sigma_{12} \end{Bmatrix} \quad (1)$$

where $\varepsilon_{11}$, $\varepsilon_{22}$ and $\varepsilon_{33}$ are the normal strains along 1, 2 and 3 directions, respectively; $\varepsilon_{12}$ is the in-plane shear strain; $\varepsilon_{13}$ and $\varepsilon_{23}$ are the transverse shear strains; $E_1$, $E_2$ and $E_3$ are Young's moduli along the 1, 2 and 3 directions, respectively; $G_{12}$, $G_{23}$ and $G_{13}$ are the shear moduli; $v_{ij}$ (i, j = 1, 2, 3) are the different Poisson's ratios.

Equation (1) can be written in another from as follows

$$\begin{Bmatrix} \varepsilon_{11} \\ \varepsilon_{22} \\ \varepsilon_{33} \\ \varepsilon_{23} \\ \varepsilon_{13} \\ \varepsilon_{12} \end{Bmatrix} = \begin{bmatrix} S_{11} & S_{12} & S_{13} & 0 & 0 & 0 \\ S_{21} & S_{22} & S_{23} & 0 & 0 & 0 \\ S_{31} & S_{32} & S_{33} & 0 & 0 & 0 \\ 0 & 0 & 0 & S_{44} & 0 & 0 \\ 0 & 0 & 0 & 0 & S_{55} & 0 \\ 0 & 0 & 0 & 0 & 0 & S_{66} \end{bmatrix} \begin{Bmatrix} \sigma_{11} \\ \sigma_{22} \\ \sigma_{33} \\ \sigma_{23} \\ \sigma_{13} \\ \sigma_{12} \end{Bmatrix} \qquad (2)$$

The elements of compliance matrix can be obtained from Eq. (1). The inverse relation of Eq. (2) is given by

$$\begin{Bmatrix} \sigma_{11} \\ \sigma_{22} \\ \sigma_{33} \\ \sigma_{23} \\ \sigma_{13} \\ \sigma_{12} \end{Bmatrix} = \begin{bmatrix} C_{11} & C_{12} & C_{13} & 0 & 0 & 0 \\ C_{21} & C_{22} & C_{23} & 0 & 0 & 0 \\ C_{31} & C_{32} & C_{33} & 0 & 0 & 0 \\ 0 & 0 & 0 & C_{44} & 0 & 0 \\ 0 & 0 & 0 & 0 & C_{55} & 0 \\ 0 & 0 & 0 & 0 & 0 & C_{66} \end{bmatrix} \begin{Bmatrix} \varepsilon_{11} \\ \varepsilon_{22} \\ \varepsilon_{33} \\ \varepsilon_{23} \\ \varepsilon_{13} \\ \varepsilon_{12} \end{Bmatrix} \qquad (3)$$

In compact form, Eqs. (2) and (3) are written as follows:

$$\{\varepsilon\} = [S]\{\sigma\} \quad \text{and} \quad \{\sigma\} = [C]\{\varepsilon\} \qquad (4)$$

where $\{\sigma\}$ and $\{\varepsilon\}$ are the stress and stain vectors, respectively. The inverse of the compliance matrix [S] is called the stiffness matrix [C].

It should be noted that the elastic constants appearing in Eq. (1) are not all independent. This is clear since the compliance matrix is symmetric. Therefore, we have the following equations relating the elastic constants:

$$\frac{v_{12}}{E_1} = \frac{v_{21}}{E_2}, \qquad \frac{v_{13}}{E_1} = \frac{v_{31}}{E_3} \quad \text{and} \quad \frac{v_{23}}{E_2} = \frac{v_{32}}{E_3} \qquad (5)$$

The above equations are called the reciprocity relations. It should be noted that the reciprocity relations can be derived irrespective of the symmetry of compliance matrix. Thus, there are nine independent elastic constants for an orthotropic material.

A material is called the transversely isotropic if its behavior in the 2-direction is identical to its behavior in the 3-direction. For this case: $E_2 = E_3$, $v_{12} = v_{13}$ and $G_{12} = G_{13}$.

When composite material is heated or cooled, the material expands or contracts. This is deformation that takes place independently of any applied load. Let $\Delta T$ be the change in temperature and let $\alpha_1$, $\alpha_2$ and $\alpha_3$ be the respective coefficients of thermal expansion for the composite material in the 1, 2 and 3 directions. In such case, the stress-strain relations can be written as follows:

$$\begin{Bmatrix} \varepsilon_{11} - \alpha_1 \Delta T \\ \varepsilon_{22} - \alpha_2 \Delta T \\ \varepsilon_{33} - \alpha_3 \Delta T \\ \varepsilon_{23} \\ \varepsilon_{13} \\ \varepsilon_{12} \end{Bmatrix} = \begin{bmatrix} S_{11} & S_{12} & S_{13} & 0 & 0 & 0 \\ S_{21} & S_{22} & S_{23} & 0 & 0 & 0 \\ S_{31} & S_{32} & S_{33} & 0 & 0 & 0 \\ 0 & 0 & 0 & S_{44} & 0 & 0 \\ 0 & 0 & 0 & 0 & S_{55} & 0 \\ 0 & 0 & 0 & 0 & 0 & S_{66} \end{bmatrix} \begin{Bmatrix} \sigma_{11} \\ \sigma_{22} \\ \sigma_{33} \\ \sigma_{23} \\ \sigma_{13} \\ \sigma_{12} \end{Bmatrix} \quad (6)$$

$$\begin{Bmatrix} \sigma_{11} \\ \sigma_{22} \\ \sigma_{33} \\ \sigma_{23} \\ \sigma_{13} \\ \sigma_{12} \end{Bmatrix} = \begin{bmatrix} C_{11} & C_{12} & C_{13} & 0 & 0 & 0 \\ C_{21} & C_{22} & C_{23} & 0 & 0 & 0 \\ C_{31} & C_{32} & C_{33} & 0 & 0 & 0 \\ 0 & 0 & 0 & C_{44} & 0 & 0 \\ 0 & 0 & 0 & 0 & C_{55} & 0 \\ 0 & 0 & 0 & 0 & 0 & C_{66} \end{bmatrix} \begin{Bmatrix} \varepsilon_{11} - \alpha_1 \Delta T \\ \varepsilon_{22} - \alpha_2 \Delta T \\ \varepsilon_{33} - \alpha_3 \Delta T \\ \varepsilon_{23} \\ \varepsilon_{13} \\ \varepsilon_{12} \end{Bmatrix} \quad (7)$$

**BASIC ASSUMPTIONS**

The purpose of this article is to predict the thermomechanical properties of a composite material by studying the micromechanics of the problem. Therefore, in reviewing models, the current study imposes the general requirements that each model must incorporate the effects of properties of constituents and their volume fractions, fiber geometry, fiber-matrix interphase, to determine a complete set of properties for the composite. In this regard, we have the following assumptions:

- both the matrix and fibers are linearly elastic, and the fibers are either isotropic or transversely isotropic,
- the fibers are continuous, identical in shape and size, and possess uniform strength,
- the fibers are spaced periodically in square- or hexagonal-packed arrays (except randomly arranged case),
- the interfacial bond between the neighbouring phases is perfect, and remains that way during deformation. Thus, the fiber-interphase-matrix debonding or matrix microcracking is not considered, and
- the resulting composite is free of voids.

Any model not meeting above-mentioned criteria is not considered here. Unless otherwise mentioned, the axis of symmetry of the reinforcement is considered to be aligned along the 1-axis. When the fibers are uniformly spaced in the matrix as shown Fig. 1, it is reasonable to consider a single fiber and the surrounding matrix material as the representative volume element (RVE).

## PRELIMINARIES

### Average stress and strain

When the composite material is loaded, the point wise stress field $\{\sigma\}(x)$ and the corresponding strain field $\{\varepsilon\}(x)$ will be non-uniform on the microscale. The solution of these non-uniform fields is a formidable problem. However, many useful results can be obtained in terms of average stress and strain [33].

Consider a RVE (V) large enough to contain many fibers, but small compared to any length scale over which the average loading or deformation of the composite varies. The volume average stress $\{\bar{\sigma}\}$ and the volume average strain $\{\bar{\varepsilon}\}$ are defined as the averages of the point wise stress $\{\sigma\}(x)$ and strain $\{\varepsilon\}(x)$ over the volume, respectively; as follows:

$$\{\bar{\sigma}\} = \frac{1}{V}\int_V \{\sigma\}(x)\, dV \quad \text{and} \quad \{\bar{\varepsilon}\} = \frac{1}{V}\int_V \{\varepsilon\}(x)\, dV \tag{8}$$

It is also convenient to define the volume average stresses and strains for the fiber and matrix phases. To obtain these, consider the volume (V) which is occupied by the volume of fibers ($V^f$) and the volume of matrix ($V^m$). In case of two-phase composite we can write

$$v_f + v_m = 1 \tag{9}$$

in which $v_f$ and $v_m$ represent the volume fractions of the fiber phase and matrix phase in the composite, respectively.

The average fiber and matrix stresses are the averages over the corresponding volumes and are given by

$$\{\bar{\sigma}^f\} = \frac{1}{V^f}\int_{V_f} \{\sigma\}(x)\, dV \quad \text{and} \quad \{\bar{\sigma}^m\} = \frac{1}{V^m}\int_{V_m} \{\sigma\}(x)\, dV \tag{10}$$

The average strains for the fiber and matrix are defined similarly.

The relationships between the fiber and matrix averages, and the overall averages can be derived from the preceding definitions and these are

$$\{\bar{\sigma}\} = v_f\{\bar{\sigma}^f\} + v_m\{\bar{\sigma}^m\} \tag{11}$$

$$\{\bar{\varepsilon}\} = v_f\{\bar{\varepsilon}^f\} + v_m\{\bar{\varepsilon}^m\} \tag{12}$$

An important related result is the average strain theorem. Let the average volume (V) subjected to the surface displacements $\{u^0\}(x)$ consistent with the uniform strain $\{\varepsilon^0\}$. Then the average strain within the region is

$$\{\bar{\varepsilon}\} = \{\varepsilon^0\} \tag{13}$$

This theorem is proved by Hill [33] by substituting the definition of the strain tensor $\{\varepsilon\}$ in terms of the displacement vector $\{u\}$ into the definition of the average strain $\{\bar{\varepsilon}\}$ and applying Gauss's theorem. The result is

$$\{\bar{\varepsilon}_{ij}\} = \frac{1}{V} \int_S \left(u_i^0 n_j + n_i u_j^0\right) dS \tag{14}$$

where S denotes the surface of V and $\{n\}$ is a unit vector normal to dS. The average strain within the volume V is completely determined by the displacements on the surface of the volume, so the displacements consistent with the uniform strain must produce the identical value of the average strain. A corollary of this principle is that, if we define a perturbation strain $\{\varepsilon^{per}\}(x)$ as the difference between the local strain and average strain as follows

$$\{\varepsilon^{per}\}(x) = \{\varepsilon\}(x) - \{\bar{\varepsilon}\} \tag{15}$$

then the volume average of $\{\varepsilon^{per}\}(x)$ must equal to zero

$$\{\bar{\varepsilon}^{per}\} = \frac{1}{V} \int_V \{\varepsilon^{per}\}(x) \, dV = 0 \tag{16}$$

The corresponding theorem for the average stress also holds. Thus, if the surface tractions consistent with $\{\sigma^0\}$ are exerted on S then the average stress is

$$\{\bar{\sigma}\} = \{\sigma^0\} \tag{17}$$

**Average properties and strain concentration**

The objective of the micromechanics models is to predict the average properties of the composite, but even these need careful definition. Here, the direct approach by Hashin [34] was followed i.e., the RVE is subjected to the surface displacements consistent with the uniform strain $\{\varepsilon^0\}$. The average stiffness of the composite is the tensor [C] that maps this uniform strain to the average stress:

$$\{\overline{\sigma}\} = [C]\{\overline{\varepsilon}\} \tag{18}$$

The average compliance matrix [S] is defined in the same way, applying tractions consistent with the uniform stress $\{\sigma^0\}$ on the surface of the average volume:

$$\{\overline{\varepsilon}\} = [S]\{\overline{\sigma}\} \tag{19}$$

Hill [33] introduced an important related concepts of strain and stress concentrations tensors and these are denoted by [M] and [N], respectively. These are essentially the ratios between the average fiber strain (or stress) and the corresponding average strain (or stress) in the composite and are given by

$$\{\overline{\varepsilon}^f\} = [M]\{\overline{\varepsilon}\} \tag{20}$$

$$\{\overline{\sigma}^f\} = [N]\{\overline{\sigma}\} \tag{21}$$

where [M] and [N] are the fourth order tensors and, in general, they must be found from a solution of the microscopic stress or strain fields. Different micromechanics models provide different ways to approximate [M] or [N]. Note that [M] and [N] have both the minor symmetries of the stiffness or the compliance tensor, but lack the major symmetry. That is,

$$M_{ijkl} = M_{jikl} = M_{ijlk} \tag{22}$$

but in general,

$$M_{ijkl} \neq M_{klij} \tag{23}$$

For later use it will be convenient to have an alternate strain concentration tensor $[\hat{M}]$ that relates the average fiber strain to the average matrix strain,

$$\{\overline{\varepsilon}^f\} = [\hat{M}]\{\varepsilon^m\} \tag{24}$$

This is related to [M] by

$$[M] = [\hat{M}]\left[(1 - v_f)[I] + v_f[\hat{M}]\right]^{-1} \tag{25}$$

in which [I] represents the fourth order unit tensor. Using equations now in hand, one can express the average composite stiffness in terms of the strain concentration tensor [M] and the fiber and matrix elastic properties [33]. Combining Eqs. (3), (14), (15), (21) and (23), we can obtain

$$[C] = [C^m] + v_f([C^f] - [C^m])[M] \tag{26}$$

The dual equation for the compliance is

$$[S] = [S^m] + v_f([S^f] - [S^m])[N] \tag{27}$$

Eqs. (26) and (27) are not independent, $[S] = [C]^{-1}$. Hence, the strain concentration tensor $[M]$ and stress concentration tensor $[N]$ are not independent either. The choice of which one to use in any instance is a matter of convenience.

To illustrate the use of the strain and stress concentration tensors, we note that the Voigt average corresponds to the assumption that the fiber and matrix both experience the same uniform strain. Then $\{\bar{\varepsilon}^f\} = \{\bar{\varepsilon}\}$, $[M] = [I]$, and from Eq. (26) the stiffness matrix of the composite is determined as

$$[C^{Voigt}] = v_f[C^f] + v_m[C^m] \tag{28}$$

The Voigt average is an upper bound on the composite stiffness. The Reuss average assumes that the fiber and matrix both experience the same uniform stress. This implies that the stress concentration tensor $[N]$ equals the unit tensor $[I]$, and from Eq. (27) the compliance matrix is given by

$$[S^{Reuss}] = v_f[S^f] + v_m[S^m] \tag{29}$$

**Eigenstrains and eigenstress**

'Eigenstrain' is a generic name given by Mura [35] to nonelastic strains such as thermal expansion, phase transformation, initial strains, plastic, misfit strains. 'Eigenstress' is a generic name given to self-equilibrated stresses caused by one or several of these eigenstrains in bodies which are free from any other external force and surface constraint. The eigenstress fields are created by the incompatibility of the eigenstrains.

For infinitesimal deformations, the total strain $\{\varepsilon_T\}$ is regarded as the sum of the elastic strain $\{\varepsilon\}$ and eigenstrain $\{\varepsilon^*\}$,

$$\{\varepsilon_T\} = \{\varepsilon\} + \{\varepsilon^*\} \tag{30}$$

The elastic strain is related to the stress by Hook's law as follows

$${\sigma} = [C]({\varepsilon} - {\varepsilon^*}) \tag{31}$$

## EFFECTIVE THERMOELASTIC PROPERTIES

This section presents the established micromechanics models for estimating the effective thermoelastic properties of a composite material containing isotropic or transversely isotropic constituents.

### Halpin-Tsai model (for isotropic constituents)

It is well known fact that the predictions for transverse Young's modulus and in-plane shear modulus using strength of materials approach do not agree well with the experimental results. This establishes a need for better modeling techniques. These techniques include numerical methods such as finite element method, boundary element method, elasticity solution, and variational principal models. Unfortunately, these models are available only as complicated equations or in graphical form [11]. Due to these difficulties, semi-empirical models have been developed for design purposes. The most useful of these models include those of Halphin and Tsai [36] because they can be used over a wide range of elastic properties and fiber volume fractions. Halphin and Tsai developed their models as simple equations by curve fitting to results that are based on elasticity. The equations are semi-empirical in nature because involved parameters in the curve fitting carry physical meaning.

The Halpin-Tsai equation for the axial Young's modulus ($E_1$) is same as that obtained through the strength of materials approach,

$$E_1 = E^f v_f + E^m v_m \tag{32}$$

Transverse Young's modulus ($E_2$):

$$\frac{E_2}{E^m} = \frac{1 + \xi \eta v_f}{1 - \eta v_f} \tag{33}$$

where

$$\eta = \frac{(E^f/E^m) - 1}{(E^f/E^m) + \xi} \tag{34}$$

The term $\xi$ is called the reinforcing factor and it depends on the fiber geometry, packing geometry and loading conditions. Halpin and Tsai obtained the value of the reinforcing factor by comparing Eqs. (33) and (34) to the solutions obtained from the elasticity solutions. For example, for circular fibers in a packing geometry of a square array, ξ = 2. For a rectangular fiber cross section of length a and width b in a hexagonal array, ξ = 2(a/b), where b is in the direction of loading.

Major Poisson's ratio ($v_{12}$):

$$v_{12} = v^f v_f + v^m v_m \tag{35}$$

In-plane shear modulus ($G_{12}$):

$$\frac{G_{12}}{G^m} = \frac{1 + \xi \eta v_f}{1 - \eta v_f} \tag{36}$$

where

$$\eta = \frac{(G^f/G^m) - 1}{(G^f/G^m) + \xi} \tag{37}$$

In this case, for circular fibers in a square array, ξ = 1. For a rectangular fiber cross sectional area of length a and width b in a hexagonal array, $\xi = \sqrt{3} \log_e(a/b)$, where a is the direction of loading.

**Strength of materials approach (for isotropic matrix)**

This section presents the strength of materials approach to determine the effective thermoelastic properties of a composite material made of transversely isotropic fiber and isotropic matrix. Using the strength of materials approach and simple rule-of-mixtures, we have the following relations (for a detailed derivation of these equations the reader is referred to Hyer [37]):

Axial Young's modulus ($E_1$):

$$E_1 = E_1^f v_f + E^m v_m \tag{38}$$

Transverse Young's modulus ($E_2$):

$$\frac{1}{E_2} = \frac{v_f}{E_2^f} + \frac{v_m}{E_m} \tag{39}$$

Major Poisson's ratio ($v_{12}$):

$$v_{12} = v^f v_f + v^m v_m \tag{40}$$

In-plane shear modulus ($G_{12}$):

$$\frac{1}{G_{12}} = \frac{v_f}{G_{12}^f} + \frac{v_m}{G^m} \tag{41}$$

Axial coefficient of thermal expansion (CTE) ($\alpha_1$):

$$\alpha_1 = \frac{\alpha_1^f E_1^f v_f + \alpha^m E^m v_m}{E_1^f v_f + E^m v_m} \tag{42}$$

Transverse CTE ($\alpha_2$):

$$\alpha_2 = \alpha_3 = \left[\alpha_2^f - \left(\frac{E^m}{E_1}\right)v_{12}^f(\alpha^m - \alpha_1^f)v_m\right]v_f + \left[\alpha^m + \left(\frac{E_1^f}{E_1}\right)v^m(\alpha^m - \alpha_1^f)v_f\right]v_m \tag{43}$$

We can also use the following rule-of-mixture relation to determine $\alpha_2$:

$$\alpha_2 = \alpha_3 = \alpha_2^f v_f + \alpha^m v_m \tag{44}$$

**Two-phase mechanics of materials approach**

This section presents the derivation of the micromechanics model using the mechanics of materials (MOM) approach [21] to determine the orthotropic effective elastic properties of a composite material made of transversely isotropic constituents. In order to satisfy the perfectly bonding situation between the fiber and matrix, researchers mainly imposed the iso-field conditions and used the rules-of-mixture [38, 39]. According to the iso-field conditions one may assume that the normal strains in the homogenized composite, fiber and matrix are equal along the fiber direction while the transverse stresses in the same phases are equal along the direction transverse to the fiber length. The rules-of-mixture allows one to express the normal stress along the fiber direction and transverse strains along the normal to the fiber direction of the homogenized composite in terms of that in the fiber and matrix and their volume fractions. Such iso-field conditions and rules-of-mixture for satisfying the perfect bonding conditions between the fiber and matrix can be expressed as

$$\begin{Bmatrix} \varepsilon_{11}^f \\ \sigma_{22}^f \\ \sigma_{33}^f \\ \sigma_{23}^f \\ \sigma_{13}^f \\ \sigma_{12}^f \end{Bmatrix} = \begin{Bmatrix} \varepsilon_{11}^m \\ \sigma_{22}^m \\ \sigma_{33}^m \\ \sigma_{23}^m \\ \sigma_{13}^m \\ \sigma_{12}^m \end{Bmatrix} = \begin{Bmatrix} \varepsilon_{11} \\ \sigma_{22} \\ \sigma_{33} \\ \sigma_{23} \\ \sigma_{13} \\ \sigma_{12} \end{Bmatrix} \tag{45}$$

$$\text{and } v_f \begin{Bmatrix} \sigma_{11}^f \\ \varepsilon_{22}^f \\ \varepsilon_{33}^f \\ \varepsilon_{23}^f \\ \varepsilon_{13}^f \\ \varepsilon_{12}^f \end{Bmatrix} + v_m \begin{Bmatrix} \sigma_{11}^m \\ \varepsilon_{22}^m \\ \varepsilon_{33}^m \\ \varepsilon_{23}^m \\ \varepsilon_{13}^m \\ \varepsilon_{12}^m \end{Bmatrix} = \begin{Bmatrix} \sigma_{11} \\ \varepsilon_{22} \\ \varepsilon_{33} \\ \varepsilon_{23} \\ \varepsilon_{13} \\ \varepsilon_{12} \end{Bmatrix} \tag{46}$$

In Eq. (46), $v_f$ is the volume fraction of the fiber with respect to the RVE volume (see Fig. 2) and $v_m = 1 - v_f$.

Substituting Eq. (3) into Eqs. (45) and (46), the stress and strain vectors in the composite material can be expressed in terms of the corresponding stress and strain vectors of the constituent phases as follows:

$$\{\sigma\} = [C_1]\{\varepsilon^f\} + [C_2]\{\varepsilon^m\} \tag{47}$$

$$\{\varepsilon\} = [V_1]\{\varepsilon^f\} + [V_2]\{\varepsilon^m\} \tag{48}$$

Also, the relations among the stresses and strains in the fiber and matrix phase given by Eq. (45) can be written as

$$[C_3]\{\varepsilon^f\} - [C_4]\{\varepsilon^m\} = 0 \tag{49}$$

The various matrices appearing in Eqs. (47)−(49) are given below

$$[C_1] = v_f \begin{bmatrix} C_{11}^f & C_{12}^f & C_{13}^f & 0 & 0 & 0 \\ 0 & 0 & 0 & 0 & 0 & 0 \\ 0 & 0 & 0 & 0 & 0 & 0 \\ 0 & 0 & 0 & 0 & 0 & 0 \\ 0 & 0 & 0 & 0 & 0 & 0 \\ 0 & 0 & 0 & 0 & 0 & 0 \end{bmatrix},$$

$$[C_2] = \begin{bmatrix} v_m C_{11}^m & v_m C_{12}^m & v_m C_{12}^m & 0 & 0 & 0 \\ C_{12}^m & C_{11}^m & C_{12}^m & 0 & 0 & 0 \\ C_{12}^m & C_{12}^m & C_{11}^m & 0 & 0 & 0 \\ 0 & 0 & 0 & C_{44}^m & 0 & 0 \\ 0 & 0 & 0 & 0 & C_{44}^m & 0 \\ 0 & 0 & 0 & 0 & 0 & C_{44}^m \end{bmatrix},$$

$$[C_3] = \begin{bmatrix} 1 & 0 & 0 & 0 & 0 & 0 \\ C^f_{12} & C^f_{22} & C^f_{23} & 0 & 0 & 0 \\ C^f_{13} & C^f_{23} & C^f_{33} & 0 & 0 & 0 \\ 0 & 0 & 0 & C^f_{44} & 0 & 0 \\ 0 & 0 & 0 & 0 & C^f_{55} & 0 \\ 0 & 0 & 0 & 0 & 0 & C^f_{66} \end{bmatrix},$$

$$[C_4] = \begin{bmatrix} 1 & 0 & 0 & 0 & 0 & 0 \\ C^m_{12} & C^m_{11} & C^m_{12} & 0 & 0 & 0 \\ C^m_{12} & C^m_{12} & C^m_{11} & 0 & 0 & 0 \\ 0 & 0 & 0 & C^m_{44} & 0 & 0 \\ 0 & 0 & 0 & 0 & C^m_{44} & 0 \\ 0 & 0 & 0 & 0 & 0 & C^m_{44} \end{bmatrix},$$

$$[V_1] = \begin{bmatrix} 0 & 0 & 0 & 0 & 0 & 0 \\ 0 & v_f & 0 & 0 & 0 & 0 \\ 0 & 0 & v_f & 0 & 0 & 0 \\ 0 & 0 & 0 & v_f & 0 & 0 \\ 0 & 0 & 0 & 0 & v_f & 0 \\ 0 & 0 & 0 & 0 & 0 & v_f \end{bmatrix}, [V_2] = \begin{bmatrix} 1 & 0 & 0 & 0 & 0 & 0 \\ 0 & v_m & 0 & 0 & 0 & 0 \\ 0 & 0 & v_m & 0 & 0 & 0 \\ 0 & 0 & 0 & v_m & 0 & 0 \\ 0 & 0 & 0 & 0 & v_m & 0 \\ 0 & 0 & 0 & 0 & 0 & v_m \end{bmatrix},$$

Using Eqs. (48) and (49), the local strain vectors $\{\varepsilon^f\}$ and $\{\varepsilon^m\}$ can be expressed in terms of the composite strain $\{\varepsilon\}$ and subsequently, using them in Eq. (47), the following constitutive relation between the state of stress and state of strain at any point in the composite material can be obtained:

$$\{\sigma\} = [C]\{\varepsilon\} \quad (50)$$

in which

$$[C] = [C_1][V_3]^{-1} + [C_2][V_4]^{-1} \quad (51)$$

$$[V_3] = [V_1] + [V_2][C_4]^{-1}[C_3] \quad \text{and} \quad [V_4] = [V_2] + [V_1][C_3]^{-1}[C_4] \quad (52)$$

**Three-phase mechanics of materials approach**

In order to realize the optimum mechanical performance of a composite material, each of its constituents must be fully utilized. This is only possible when a certain interaction exists between the fiber and matrix. A method for improving the fiber-matrix interaction is the surface treatment of the fibers before they are impregnated in the matrix. Most of the glass fibers used for filament winding, knitted and nonknitted woven fabrics undergo a single and final treatment just in the process of their formation [40]. Such treatment is called sizing. The treating solution is a mixture of many components, with the coupling agent, binder, and lubricant. By applying different types or different levels of treatment, it is possible to improve the quality of the bond interaction between the fiber and matrix. Such an interphase can be assumed an intermediate phase between the fiber and matrix, and based on the principal material coordinate (1–2–3) axes shown in Fig. 3, the constitutive relations for the constituent phases of the three-phase composite material are written as

$$\{\sigma^r\} = [C^r]\{\varepsilon^r\}; \quad r = f, i \text{ and } m \tag{53}$$

In Eq. (53), the superscripts f, i and m denote the fiber, interphase and matrix, respectively. Iso-field conditions and rules-of-mixture [22] for satisfying the perfect bonding conditions between the fiber and neighboring phases can be expressed as

$$\begin{Bmatrix} \varepsilon_{11}^f \\ \sigma_{22}^f \\ \sigma_{33}^f \\ \sigma_{23}^f \\ \sigma_{13}^f \\ \sigma_{12}^f \end{Bmatrix} = \begin{Bmatrix} \varepsilon_{11}^i \\ \sigma_{22}^i \\ \sigma_{33}^i \\ \sigma_{23}^i \\ \sigma_{13}^i \\ \sigma_{12}^i \end{Bmatrix} = \begin{Bmatrix} \varepsilon_{11}^m \\ \sigma_{22}^m \\ \sigma_{33}^m \\ \sigma_{23}^m \\ \sigma_{13}^m \\ \sigma_{12}^m \end{Bmatrix} = \begin{Bmatrix} \varepsilon_{11} \\ \sigma_{22} \\ \sigma_{33} \\ \sigma_{23} \\ \sigma_{13} \\ \sigma_{12} \end{Bmatrix} \tag{54}$$

and

$$v_f \begin{Bmatrix} \sigma_{11}^f \\ \varepsilon_{22}^f \\ \varepsilon_{33}^f \\ \varepsilon_{23}^f \\ \varepsilon_{13}^f \\ \varepsilon_{12}^f \end{Bmatrix} + v_i \begin{Bmatrix} \sigma_{11}^i \\ \varepsilon_{22}^i \\ \varepsilon_{33}^i \\ \varepsilon_{23}^i \\ \varepsilon_{13}^i \\ \varepsilon_{12}^i \end{Bmatrix} + v_m \begin{Bmatrix} \sigma_{33}^m \\ \varepsilon_{22}^m \\ \varepsilon_{33}^m \\ \varepsilon_{23}^m \\ \varepsilon_{13}^m \\ \varepsilon_{12}^m \end{Bmatrix} = \begin{Bmatrix} \sigma_{11} \\ \varepsilon_{22} \\ \varepsilon_{33} \\ \varepsilon_{23} \\ \varepsilon_{13} \\ \varepsilon_{12} \end{Bmatrix} \tag{55}$$

In Eq. (55), $v_f$, $v_i$ and $v_m$ represent the volume fractions of the fiber, interphase and matrix, respectively, present in the RVE. Substituting Eqs. (3) into Eqs. (54) and (55), the stress and strain vectors in the composite material can be expressed in terms of the corresponding stress and strain vectors of the constituent phases as follows:

$$\{\sigma\} = [C_1]\{\varepsilon^f\} + [C_2]\{\varepsilon^i\} + [C_3]\{\varepsilon^m\} \tag{56}$$

$$\{\varepsilon\} = [V_1]\{\varepsilon^f\} + [V_2]\{\varepsilon^i\} + [V_3]\{\varepsilon^m\} \tag{57}$$

$$[C_4]\{\varepsilon^f\} - [C_5]\{\varepsilon^i\} = 0 \qquad (58)$$

$$[C_5]\{\varepsilon^i\} - [C_6]\{\varepsilon^m\} = 0 \qquad (59)$$

The various matrices appearing in Eqs. (56)–(59) are given by

$$[C_1] = \begin{bmatrix} v_f C_{11}^f & v_f C_{12}^f & v_f C_{13}^f & 0 & 0 & 0 \\ C_{12}^f & C_{22}^f & C_{23}^f & 0 & 0 & 0 \\ C_{13}^f & C_{23}^f & C_{33}^f & 0 & 0 & 0 \\ 0 & 0 & 0 & C_{44}^f & 0 & 0 \\ 0 & 0 & 0 & 0 & C_{55}^f & 0 \\ 0 & 0 & 0 & 0 & 0 & C_{66}^f \end{bmatrix}, \quad [C_2] = v_i \begin{bmatrix} C_{11}^i & C_{12}^i & C_{13}^i & 0 & 0 & 0 \\ 0 & 0 & 0 & 0 & 0 & 0 \\ 0 & 0 & 0 & 0 & 0 & 0 \\ 0 & 0 & 0 & 0 & 0 & 0 \\ 0 & 0 & 0 & 0 & 0 & 0 \\ 0 & 0 & 0 & 0 & 0 & 0 \end{bmatrix},$$

$$[C_3] = v_m \begin{bmatrix} C_{11}^m & C_{12}^m & C_{13}^m & 0 & 0 & 0 \\ 0 & 0 & 0 & 0 & 0 & 0 \\ 0 & 0 & 0 & 0 & 0 & 0 \\ 0 & 0 & 0 & 0 & 0 & 0 \\ 0 & 0 & 0 & 0 & 0 & 0 \\ 0 & 0 & 0 & 0 & 0 & 0 \end{bmatrix}, \quad [C_4] = \begin{bmatrix} 1 & 0 & 0 & 0 & 0 & 0 \\ C_{12}^f & C_{22}^f & C_{23}^f & 0 & 0 & 0 \\ C_{13}^f & C_{23}^f & C_{33}^f & 0 & 0 & 0 \\ 0 & 0 & 0 & C_{44}^f & 0 & 0 \\ 0 & 0 & 0 & 0 & C_{55}^f & 0 \\ 0 & 0 & 0 & 0 & 0 & C_{66}^f \end{bmatrix},$$

$$[C_5] = \begin{bmatrix} 1 & 0 & 0 & 0 & 0 & 0 \\ C_{12}^i & C_{22}^i & C_{23}^i & 0 & 0 & 0 \\ C_{13}^i & C_{23}^i & C_{33}^i & 0 & 0 & 0 \\ 0 & 0 & 0 & C_{44}^i & 0 & 0 \\ 0 & 0 & 0 & 0 & C_{55}^i & 0 \\ 0 & 0 & 0 & 0 & 0 & C_{66}^i \end{bmatrix}, \quad [C_6] = \begin{bmatrix} 1 & 0 & 0 & 0 & 0 & 0 \\ C_{12}^m & C_{22}^m & C_{23}^m & 0 & 0 & 0 \\ C_{13}^m & C_{23}^m & C_{33}^m & 0 & 0 & 0 \\ 0 & 0 & 0 & C_{44}^m & 0 & 0 \\ 0 & 0 & 0 & 0 & C_{55}^m & 0 \\ 0 & 0 & 0 & 0 & 0 & C_{66}^m \end{bmatrix},$$

$$[V_1] = \begin{bmatrix} 1 & 0 & 0 & 0 & 0 & 0 \\ 0 & v_f & 0 & 0 & 0 & 0 \\ 0 & 0 & v_f & 0 & 0 & 0 \\ 0 & 0 & 0 & v_f & 0 & 0 \\ 0 & 0 & 0 & 0 & v_f & 0 \\ 0 & 0 & 0 & 0 & 0 & v_f \end{bmatrix}, \quad [V_2] = \begin{bmatrix} 0 & 0 & 0 & 0 & 0 & 0 \\ 0 & v_i & 0 & 0 & 0 & 0 \\ 0 & 0 & v_i & 0 & 0 & 0 \\ 0 & 0 & 0 & v_i & 0 & 0 \\ 0 & 0 & 0 & 0 & v_i & 0 \\ 0 & 0 & 0 & 0 & 0 & v_i \end{bmatrix}, \quad \text{and}$$

$$[V_3] = \begin{bmatrix} 0 & 0 & 0 & 0 & 0 & 0 \\ 0 & v_m & 0 & 0 & 0 & 0 \\ 0 & 0 & v_m & 0 & 0 & 0 \\ 0 & 0 & 0 & v_m & 0 & 0 \\ 0 & 0 & 0 & 0 & v_m & 0 \\ 0 & 0 & 0 & 0 & 0 & v_m \end{bmatrix} \qquad (60)$$

Using Eqs. (57)–(59), the local strain vectors can be expressed in terms of the composite strain and subsequently, using them in Eq. (56), the following effective elastic coefficient matrix of the composite can be obtained:

$$[C] = [C_1][V_5]^{-1} + [C_7][V_6]^{-1} \tag{61}$$

and

$$[C_7] = [C_3] + [C_2][C_5]^{-1}[C_6], \quad [V_4] = [V_3] + [V_2][C_5]^{-1}[C_6],$$

$$[V_5] = [V_1] + [V_4][C_6]^{-1}[C_4] \text{ and } [V_6] = [V_4] + [V_1][C_4]^{-1}[C_6] \tag{62}$$

**Two-phase Mori-Tanaka method**

Mori-Tanaka (MT) method [12] is an Eshelby-type model [41] which accounts for interaction among the neighboring reinforcements. Due to its simplicity, the MT model has been reported to be the efficient analytical model for predicting the effective orthotropic elastic properties of composites. According to the two-phase MT method, the effective coefficient matrix [C] of the composite is given by [42]:

$$[C] = [C^m] + v_f([C^f] - [C^m])[A_1] \tag{63}$$

in which

$$[A_1] = \langle[A]\rangle\left[v_m[I] + v_f[\tilde{A}_1]\right]^{-1} \text{ and } \langle[A]\rangle = \left[[I] + [S]([C^m])^{-1}([C^f] - [C^m])\right]^{-1} \tag{64}$$

where $[S_f]$ is the Eshelby tensor and it is computed based on the properties of the matrix and shape of the fiber. The elements of the Eshelby tensor for the cylindrical reinforcement in the isotropic matrix are explicitly given by [13]:

$$[S] = \begin{bmatrix} S_{1111} & S_{1122} & S_{1133} & 0 & 0 & 0 \\ S_{2211} & S_{2222} & S_{2233} & 0 & 0 & 0 \\ S_{3311} & S_{3322} & S_{3333} & 0 & 0 & 0 \\ 0 & 0 & 0 & S_{2323} & 0 & 0 \\ 0 & 0 & 0 & 0 & S_{1313} & 0 \\ 0 & 0 & 0 & 0 & 0 & S_{1212} \end{bmatrix}$$

in which

$$S_{1111} = 0, \quad S_{2222} = S_{3333} = \frac{5 - 4v^m}{8(1 - v^m)}, \quad S_{2211} = S_{3311} = \frac{v^m}{2(1 - v^m)},$$

$$S_{2233} = S_{3322} = \frac{4\nu^m - 1}{8(1 - \nu^m)}, \quad S_{1122} = S_{1133} = 0, \quad S_{1313} = S_{1212} = 1/4$$

$$\text{and} \quad S_{2323} = \frac{3 - 4\nu^m}{8(1 - \nu^m)}$$

**Three-phase Mori-Tanaka method**

The effective elastic properties of the composite can be estimated in the presence of an interphase between a fiber and the matrix. Employing the procedure of the MT model for multiple inclusions [43], a three-phase MT model can be derived for the three-phase composite. The explicit formulation of such three-phase MT model can be derived as

$$[C] = \left[v_m [C^m][I] + v_f [C^f][A_f] + v_i [C^i][A_i]\right] \left[v_m [I] + v_f [A_f] + v_i [A_i]\right]^{-1} \quad (65)$$

In Eq. (65), $v_f$, $v_i$ and $v_m$ represent the volume fractions of the fiber, interphase and matrix, respectively with respect to the RVE of the composite. The concentration tensors $[A_f]$ and $[A_i]$ appearing in Eq. (65) are given by

$$[A_f] = \left[[I] + [S_f]\{([C^m])^{-1}([C^f] - [C^m])\}\right]^{-1} \quad (66)$$

$$[A_i] = \left[[I] + [S_i]\{([C^m])^{-1}([C^i] - [C^m])\}\right]^{-1} \quad (67)$$

Furthermore, in the above matrices $[S_f]$ and $[S_i]$ indicate the Eshelby tensors for the domains f and i, respectively. For the cylindrical fiber inclusion in the interphase, the elements of $[S_f]$ can be written as follows:

$$S_{1111} = 0, \quad S_{2222} = S_{3333} = \frac{5 - 4\nu^i}{8(1 - \nu^i)}, \quad S_{2211} = S_{3311} = \frac{\nu^i}{2(1 - \nu^m)},$$

$$S_{2233} = S_{3322} = \frac{4\nu^i - 1}{8(1 - \nu^i)}, \quad S_{1122} = S_{1133} = 0, \quad S_{1313} = S_{1212} = 1/4$$

$$\text{and} \quad S_{2323} = \frac{3 - 4\nu^i}{8(1 - \nu^i)}$$

while for the cylindrical fiber-interphase inclusion in the matrix, the elements of $[S_i]$ are given as follows:

$$S_{1111} = 0, \quad S_{2222} = S_{3333} = \frac{5 - 4\nu^m}{8(1 - \nu^m)}, \quad S_{2211} = S_{3311} = \frac{\nu^m}{2(1 - \nu^m)},$$

$$S_{2233} = S_{3322} = \frac{4\nu^m - 1}{8(1 - \nu^m)}, \quad S_{1122} = S_{1133} = 0, \quad S_{1313} = S_{1212} = 1/4$$

$$\text{and} \quad S_{2323} = \frac{3 - 4\nu^m}{8(1 - \nu^m)}$$

Using the effective elastic coefficient matrix [C], the effective thermal expansion coefficient vector {α} for the composite can be determined as follows [44]:

$$\{\alpha\} = \{\alpha^f\} + \left([C]^{-1} - [C^f]^{-1}\right)\left([C^f]^{-1} - [C^m]^{-1}\right)^{-1}\left(\{\alpha^f\} - \{\alpha^m\}\right) \qquad (68)$$

where $\{\alpha^f\}$ and $\{\alpha^m\}$ are the thermal expansion coefficient vectors of the fiber and matrix, respectively.

**Composite cylindrical assemblage model**

The effective elastic moduli of a composite material reinforced with aligned hollow circular fibers were derived by Hashin and Rosen [8] using a variational method. Their model is known as composite cylindrical assemblage model. Schematic of such model is demonstrated in Fig. 4. They obtained the following elastic moduli for hexagonal arrays (see Fig 5) of identical fibers reinforced in the matrix.

Axial Young's modulus ($E_1$):

$$E_1 = E^m \left(v_f \frac{E^f}{E^m} + v_m\right) \frac{E^m(A_1 - A_3 B_1) + E^f(A_2 - A_4 B_2)}{E^m(A_1 - A_3) + E^f(A_2 - A_4)} \qquad (69)$$

$$A_1 = \frac{1 + a^2}{1 - a^2} - v_f, \quad A_2 = \frac{1 + v_g}{v_m} + v_m, \quad A_3 = \frac{2v_f^2}{1 - a^2}, \quad A_4 = \frac{2v_m^2 v_g}{v_m},$$

$$B_1 = \frac{v_m v_f E^f + v_f v_m E^m}{v_f v_f E^f + v_m v_m E^m}, \quad B_2 = \frac{v_f B_1}{v_m}, \quad a = \frac{R_1}{R_2}, \quad v_g = \frac{R_2^2}{R^2}, \quad v_f = (1 - a^2)v_g \text{ and}$$

$$v_m = 1 - v_g \qquad (70)$$

where $v_g$ is the volume fraction of the gross cylindrical inclusion (fiber and the surrounding bonding matrix; see Fig. 4); R, $R_1$ and $R_2$ are the inner radius of the hollow fiber, outer radius of the hollow fiber and radius of gross cylindrical inclusion, respectively. It may be noted that the elastic properties of the bonding matrix are same as those of the matrix.

Longitudinal Poisson's ratio ($v_{12}$):

$$v_{12} = \frac{E^f L_1 v_f + v^m E^m L_2 v_m}{E^f L_3 v_f + E^m L_2 v_m} \tag{71}$$

$$L_1 = 2v^f[1-(v^m)^2]v_g + v_m[1+v^m]v^m, \qquad L_2 = \left[(1+v^f)a^2 + 1 - v^f - 2(v^f)^2\right]v_g$$

$$\text{and} \quad L_3 = 2[1-(v^m)^2]v_g + v_m[1+v^m] \tag{72}$$

Longitudinal shear modulus ($G_{12}$):

$$G_{12} = G^m \frac{\eta(1-a^2)(1+v_g) + v_m(1+a^2)}{\eta(1-a^2)v_m + (1+a^2)(1+v_g)} \tag{73}$$

where $\eta = G^f/G^m$

Bulk modulus ($K_{23}$):

$$K_{23} = \bar{K}^m \frac{\Phi(1-a^2)(1+2v^m v_g) + 2v^m v_m \left(1+\frac{a^2}{2v^f}\right)}{\Phi(1-a^2)v_m + (v_g + 2v^m)\left(1+\frac{a^2}{2v^f}\right)} \tag{74}$$

where $\Phi = \bar{K}^f/\bar{K}^m$, $\quad \bar{K}^f = \lambda^f + G^f$, $\quad \bar{K}^m = \lambda^m + G^m$, $\quad \lambda^f = K^f - \frac{2G^f}{3}$, $\quad \lambda^m = K^m - \frac{2G^m}{3}$

in which $\lambda^f$ and $\lambda^m$ are the Lamè constants of the fiber and matrix, respectively.

Transverse shear modulus ($G_{23}$):

Given the complexity of the transverse shear modulus, the importance of demonstrating the bounding of the solution is seen. The upper bounds for $G_{23}$, given by Hashin and Rosen (1964), are

$$G_{23}^{(+)} = G^m \left[1 - \frac{2v_g(1-v^m)\bar{A}_4^\varepsilon}{1-2v^m}\right] \tag{75}$$

where $\bar{A}_4^\varepsilon$ can be obtained from the solution of the following relation

$$\begin{bmatrix} 1 & 1/v_g & v_g^2 & v_g & 0 & 0 & 0 & 0 \\ 0 & -\dfrac{3-4\nu^m}{(3-2\nu^m)v_g} & -2v_g^2 & \dfrac{v_g}{1-2\nu^m} & 0 & 0 & 0 & 0 \\ 1 & 1 & 1 & 1 & -1 & -1 & -1 & -1 \\ 0 & -\dfrac{3-4\nu^m}{3-2\nu^m} & -2 & \dfrac{1}{1-2\nu^m} & 0 & \dfrac{3-4\nu^f}{3-2\nu^f} & 2 & -\dfrac{1}{1-2\nu^f} \\ 1 & \dfrac{3}{3-2\nu^m} & -3 & \dfrac{1}{1-2\nu^m} & -\eta & -\dfrac{3\eta}{3-2\nu^f} & 3\eta & -\dfrac{\eta}{1-2\nu^f} \\ 0 & -\dfrac{1}{3-2\nu^m} & 2 & -\dfrac{1}{1-2\nu^m} & 0 & \dfrac{\eta}{3-2\nu^f} & -2\eta & \dfrac{\eta}{1-2\nu^f} \\ 0 & 0 & 0 & 0 & 1 & \dfrac{3a^2}{3-2\nu^f} & -3/a^4 & \dfrac{1}{(1-2\nu^f)a^2} \\ 0 & 0 & 0 & 0 & 0 & -\dfrac{a^2}{3-2\nu^f} & 2/a^4 & -\dfrac{1}{(1-2\nu^f)a^2} \end{bmatrix} \begin{Bmatrix} \bar{A}_1^\varepsilon \\ \bar{A}_2^\varepsilon \\ \bar{A}_3^\varepsilon \\ \bar{A}_4^\varepsilon \\ \bar{B}_1^\varepsilon \\ \bar{B}_2^\varepsilon \\ \bar{B}_3^\varepsilon \\ \bar{B}_4^\varepsilon \end{Bmatrix} = \begin{Bmatrix} 1 \\ 0 \\ 0 \\ 0 \\ 0 \\ 0 \\ 0 \\ 0 \end{Bmatrix}$$

The lower bounds for $G_{23}$, are given by [8]

$$G_{23}^{(-)} = \dfrac{G^m}{\left[1 + \dfrac{2v_g(1-\nu^m)\bar{A}_4^\sigma}{1-2\nu^m}\right]} \tag{76}$$

where $\bar{A}_4^\sigma$ can be obtained from the solution of the following relation

$$\begin{bmatrix} 1 & \dfrac{3}{(3-2\nu^m)v_g} & -3v_g^2 & \dfrac{v_g}{1-2\nu^m} & 0 & 0 & 0 & 0 \\ 0 & -\dfrac{1}{(3-2\nu^m)v_g} & 2v_g^2 & -\dfrac{v_g}{1-2\nu^m} & 0 & 0 & 0 & 0 \\ 1 & 1 & 1 & 1 & -1 & -1 & -1 & -1 \\ 0 & -\dfrac{3-4\nu^m}{3-2\nu^m} & -2 & \dfrac{1}{1-2\nu^m} & 0 & \dfrac{3-4\nu^f}{3-2\nu^f} & 2 & -\dfrac{1}{1-2\nu^f} \\ 1 & \dfrac{3}{3-2\nu^m} & -3 & \dfrac{1}{1-2\nu^m} & -\eta & -\dfrac{3\eta}{3-2\nu^f} & 3\eta & -\dfrac{\eta}{1-2\nu^f} \\ 0 & -\dfrac{1}{3-2\nu^m} & 2 & -\dfrac{1}{1-2\nu^m} & 0 & \dfrac{\eta}{3-2\nu^f} & -2\eta & \dfrac{\eta}{1-2\nu^f} \\ 0 & 0 & 0 & 0 & 1 & \dfrac{3a^2}{3-2\nu^f} & -3/a^4 & \dfrac{1}{(1-2\nu^f)a^2} \\ 0 & 0 & 0 & 0 & 0 & -\dfrac{a^2}{3-2\nu^f} & 2/a^4 & -\dfrac{1}{(1-2\nu^f)a^2} \end{bmatrix} \begin{Bmatrix} \bar{A}_1^\sigma \\ \bar{A}_2^\sigma \\ \bar{A}_3^\sigma \\ \bar{A}_4^\sigma \\ \bar{B}_1^\sigma \\ \bar{B}_2^\sigma \\ \bar{B}_3^\sigma \\ \bar{B}_4^\sigma \end{Bmatrix} = \begin{Bmatrix} 1 \\ 0 \\ 0 \\ 0 \\ 0 \\ 0 \\ 0 \\ 0 \end{Bmatrix}$$

**Method of cells approach**

This section presents the micromechanics model based on the method of cells approach to estimate the effective thermoelastic properties of a composite. Assuming that fibers are uniformly spaced in the matrix and aligned along the 1–axis, the resulting composite can be viewed to be composed of cells forming doubly periodic arrays along the 2– and 3–directions.

Such an arrangement of cells consisting of subcells is illustrated in Fig. 6. Each rectangular parallelepiped subcell is labeled by β γ, with β and γ denoting the location of the subcell along the 2–and 3–directions, respectively. The numbers of subcells present in the cell along the 2– and 3–directions are represented by M and N, respectively. Here, each cell represents the RVE and subcell can be either a fiber or the matrix. Modeling the perfectly bonding condition at the interface between the subcells is the basis for deriving the micromechanics model using the method of cells approach [14]. It may be noted that such perfectly bonding conditions between the subcells can be established by satisfying the compatibility of displacements and continuities of tractions at the interfaces between the subcells of the cell. The volume $V_{\beta\gamma}$ of each subcell is

$$V_{\beta\gamma} = lb_\beta h_\gamma \tag{77}$$

where $b_\beta$, $h_\gamma$ and l denote the width, height and length of the subcell, respectively, while the volume (V) of the cell is

$$V = lbh \tag{78}$$

with

$$b = \sum_{\beta=1}^{M} b_\beta \text{ and } h = \sum_{\gamma=1}^{N} h_\gamma$$

The constitutive relations for the medium of a subcell under thermal environment are given by

$$\{\sigma^{\beta\gamma}\} = [C^{\beta\gamma}](\{\varepsilon^{\beta\gamma}\} - \{\alpha^{\beta\gamma}\}\Delta T) \tag{79}$$

where $\{\sigma^{\beta\gamma}\}$, $\{\varepsilon^{\beta\gamma}\}$, $\{\alpha^{\beta\gamma}\}$ and $[C^{\beta\gamma}]$ represent the stress vector, strain vector, thermal expansion coefficient vector and elastic coefficient matrix of the subcell, respectively. It should be noted that in order to utilize the constituent material properties during computation, the superscript βγ denoting the location of the subcell in the cell should be replaced by f or m according as the medium of the subcell is the fiber or the matrix, respectively. In the method of cells approach, the effective thermoelastic properties are determined by evaluating the thermoelastic properties of the repeating cells filled up with the equivalent homogeneous materials. This amounts to volume averaging of the field variable in concern. Thus, the volume averaged strain and stress in the composite can be expressed as follows

$$\{\varepsilon^c\} = \frac{1}{V}\sum_{\beta=1}^{M}\sum_{\gamma=1}^{N} V_{\beta\gamma}\{\varepsilon^{\beta\gamma}\} \quad \text{and} \quad \{\sigma^c\} = \frac{1}{V}\sum_{\beta=1}^{M}\sum_{\gamma=1}^{N} V_{\beta\gamma}\{\sigma^{\beta\gamma}\} \tag{80}$$

Here, the superscript c designates the composite. Imposition of the interfacial displacement continuities provides the following $2(M + N) + MN + 1$ number of relations between the volume averaged subcell strains and the composite strains:

$$\varepsilon_{11}^{\beta\gamma} = \varepsilon_{11}^c, \qquad \beta = 1,2,\ldots M, \quad \gamma = 1,2,\ldots,N \tag{81}$$

$$\sum_{\gamma=1}^{N} h_\gamma \varepsilon_{22}^{\beta\gamma} = h\varepsilon_{22}^c, \qquad \beta = 1,2,\ldots,M \tag{82}$$

$$\sum_{\beta=1}^{M} b_\beta \varepsilon_{33}^{\beta\gamma} = b\varepsilon_{33}^c, \qquad \gamma = 1,2,\ldots,N \tag{83}$$

$$\sum_{\beta=1}^{M}\sum_{\gamma=1}^{N} b_\beta h_\gamma \varepsilon_{23}^{\beta\gamma} = bh\varepsilon_{23}^c, \tag{84}$$

$$\sum_{\gamma=1}^{N} h_\gamma \varepsilon_{12}^{\beta\gamma} = h\varepsilon_{12}^c, \qquad \beta = 1,2,\ldots,M \tag{85}$$

$$\sum_{\beta=1}^{M} b_\beta \varepsilon_{13}^{\beta\gamma} = b\varepsilon_{13}^c, \qquad \gamma = 1,2,\ldots,N \tag{86}$$

Imposition of the interfacial traction continuity conditions between the adjacent subcells yields the following $5MN - 2(M + N) - 1$ number of relations between the volume averaged subcell stresses:

$$\sigma_{22}^{\beta\gamma} = \sigma_{22}^{(\beta+1)\gamma}, \qquad \beta = 1,2,\ldots,M-1, \quad \gamma = 1,2,\ldots,N \tag{87}$$

$$\sigma_{33}^{\beta\gamma} = \sigma_{33}^{\beta(\gamma+1)}, \qquad \beta = 1,2,\ldots,M, \quad \gamma = 1,2,\ldots,N-1 \tag{88}$$

$$\sigma_{13}^{\beta\gamma} = \sigma_{13}^{(\beta+1)\gamma}, \qquad \beta = 1,2,\ldots,M-1,, \quad \gamma = 1,2,\ldots,N \tag{89}$$

$$\sigma_{12}^{\beta\gamma} = \sigma_{12}^{\beta(\gamma+1)}, \qquad \beta = 1,2,\ldots,M, \quad \gamma = 1,2,\ldots,N-1 \qquad (90)$$

$$\sigma_{23}^{\beta\gamma} = \sigma_{23}^{(\beta+1)\gamma}, \qquad \beta = 1,2,\ldots,M-1, \quad \gamma = 1,2,\ldots,N \qquad (91)$$

$$\sigma_{23}^{\beta\gamma} = \sigma_{23}^{\beta(\gamma+1)}, \qquad \beta = 1, \qquad \gamma = 1,2,\ldots,N-1 \qquad (92)$$

Equations (81)–(86) form a set of $2(M+N) + MN + 1$ number of relations and can be arranged in a matrix form as follows:

$$[A_G]\{\varepsilon_s\} = [B]\{\varepsilon^c\} \qquad (93)$$

in which $\{\varepsilon_s\}$ is the $(6MN \times 1)$ vector of subcell strains assembled together and $\{\varepsilon^c\}$ is the $(6 \times 1)$ vector of composite strains; $[A_G]$ is the $[(2(M+N) + MN + 1) \times (6MN)]$ matrix formed by the geometrical parameters of the subcells; and $[B]$ matrix is constructed by the geometrical parameters of the cell.

Using the constitutive equations given by Eq. (79), $(5MN - 2(M+N) - 1)$ number of traction continuity conditions can be expressed in the following matrix form:

$$[A_M](\{\varepsilon_s\} - \{\alpha_s\}\Delta T) = 0 \qquad (94)$$

in which $\{\alpha_s\}$ is the $(6MN \times 1)$ thermal expansion coefficient vector of subcells assembled together and $[A_M]$ is the $[(5MN - 2(M+N) - 1) \times (6MN)]$ matrix containing the elastic properties of the subcells. Combination of Eqs. (93) and (94) leads to

$$[A]\{\varepsilon_s\} = [K]\{\varepsilon^c\} + [D]\{\alpha_s\}\Delta T \qquad (95)$$

where

$$[A] = \begin{bmatrix} [A_M] \\ [A_G] \end{bmatrix}, \qquad [K] = \begin{bmatrix} \bar{0} \\ [B] \end{bmatrix} \quad \text{and} \quad [D] = \begin{bmatrix} [A_M] \\ \bar{\bar{0}} \end{bmatrix}$$

with $\bar{0}$ and $\bar{\bar{0}}$ being $[(5MN - 2(M+N) - 1) \times (6)]$ and $[(2(M+N) + MN + 1) \times (6MN)]$ null vectors, respectively. From Eq. (95), the subcell strains can be expressed in terms of the composite strains as follows:

$$\{\varepsilon_s\} = [A_c]\{\varepsilon^c\} + [D_c]\{\alpha_s\}\Delta T \qquad (96)$$

where $[A_c] = [A]^{-1}[K]$ and $[D_c] = [A]^{-1}[D]$. The matrices $[A_c]$ and $[D_c]$ can be treated as the mechanical and thermal concentration matrices, respectively. It is now possible to extract the matrices $[A_c^{\beta\gamma}]$ and $[D_c^{\beta\gamma}]$ of strain concentration factors for each subcell from the matrices $[A_c]$ and $[D_c]$, respectively, such that each subcell strains can be expressed in terms of the composite strains and thermal strains as follows:

$$\{\varepsilon^{\beta\gamma}\} = [A_c^{\beta\gamma}]\{\varepsilon^c\} + [D_c^{\beta\gamma}]\{\alpha_s\}\Delta T \tag{97}$$

Substituting Eq. (97) into Eq. (79) yields

$$\{\sigma^{\beta\gamma}\} = [C^{\beta\gamma}]([A_c^{\beta\gamma}]\{\varepsilon^c\} + [D_c^{\beta\gamma}]\{\alpha_s\}\Delta T - \{\alpha^{\beta\gamma}\}\Delta T) \tag{98}$$

Using Eq. (97) in Eq. (80), the constitutive relations for the composite can be derived as

$$\{\sigma^c\} = [C](\{\varepsilon^c\} - \{\alpha\}\Delta T) \tag{99}$$

in which the effective elastic coefficient matrix $[C]$ and effective thermal expansion coefficient vector $\{\alpha\}$ of the composite are given by

$$[C] = \frac{1}{V}\sum_{\beta=1}^{M}\sum_{\gamma=1}^{N} V_{\beta\gamma}[C^{\beta\gamma}][A_c^{\beta\gamma}] \tag{100}$$

$$\{\alpha\} = -\frac{[C]^{-1}}{V}\sum_{\beta=1}^{M}\sum_{\gamma=1}^{N} V_{\beta\gamma}[C^{\beta\gamma}]([D_c^{\beta\gamma}]\{\alpha_s\} - \{\alpha^{\beta\gamma}\}) \tag{101}$$

**Global coordinate system**

So far, in this work, the constitutive relations for any material system are written in terms of the thermoelastic coefficients that are referred to the principal material coordinate system. The coordinate system used in the problem formulation, in general, does not always coincide with the principal coordinate system. Further, composite laminates have several laminae; each with different orientation. Therefore, there is a need to establish transformation relations among the thermoelastic coefficients in one coordinate system to the corresponding properties in another coordinate system.

In forming composite laminates, fiber reinforced laminae are stacked each other but each having its own fiber direction. Figure 7 illustrate such rotation of fibers with respect to the 1-axis.

If the 3-axis of the problem is taken along the laminate thickness, we have the following transformation law to relate the thermoelastic coefficients matrices ($[\bar{C}]$ and $[\bar{\alpha}]$) in the problem coordinates (1'-2'-3') to those of $[C]$ and $\{\alpha\}$ in the material coordinates (1-2-3):

$$[\bar{C}] = [T]^{-T}[C][T] \quad \text{and} \quad [\bar{\alpha}] = [T]^{-T}\{\alpha\} \tag{102}$$

where, $[T] = \begin{bmatrix} m^2 & n^2 & 0 & 0 & 0 & 2mn \\ n^2 & m^2 & 0 & 0 & 0 & -2mn \\ 0 & 0 & 1 & 0 & 0 & 0 \\ 0 & 0 & 0 & m & n & 0 \\ 0 & 0 & 0 & n & m & 0 \\ -mn & mn & 0 & 0 & 0 & m^2 - n^2 \end{bmatrix}$ with $m = \cos\theta$ and $n = \sin\theta$.

**Composite containing randomly oriented fibers**

Among various forms of fiber reinforced composites, sheet moulding compound (SMC) appears to be one of the most promising candidates forming potential applications in automotive industry because of its relatively low cost and suitability for high-volume production. The commonly used fibers of automotive-type SMC are coated S-glass filaments. Such glass fibers are randomly dispersed in a filled or an unfilled thermosetting resin. Considering different cases, several models were provided herein to determine the elastic properties of the composite reinforced with randomly oriented fibers.

*Voigt-Reuss model*

This model estimates Young's modulus ($E_{VR}$) for a lamina of short random fibers and does not take into account any geometry of the reinforcement. The following relation provides the effective Young's modulus:

$$E_{VR} = \frac{3}{8}E_A + \frac{5}{8}E_T \tag{103}$$

where $E_A$ and $E_T$ are the effective axial and transverse Young's moduli of the composite containing aligned fibers, respectively, and are given by

$$E_A = v_f E^f + (1 - v_f)E^m \quad \text{and} \quad E_T = \frac{E^f E^m}{E^f(1 - v_f) + v_f E^m} \tag{104}$$

*Halpin-Tsai model*

This model estimates Young's modulus ($E_{HR}$) for a lamina of short random fibers considering the effect of geometrical parameters of the reinforcement.

$$E_{HT} = \frac{3}{8} E_A + \frac{5}{8} E_T \qquad (105)$$

Various relations appeared in Eq. (105) are given as follows:

$$E_A = \frac{1 + 2(L_f/d_f)\eta_A v_f E^m}{1 - \eta_A v_f}, \quad E_T = \frac{1 + 2\eta_T v_f E^m}{1 - \eta_L v_f}, \quad \eta_A = \frac{(E^f/E^m) - 1}{(E^f/E^m) + 2(L_f/d_f)} \quad \text{and}$$

$$\eta_T = \frac{(E^f/E^m) - 1}{(E^f/E^m) + 2}$$

where $d_f$ and $L_f$ denote the respective diameter and length of a fiber.

*Modified mixture law*

The formulation of Piggot for three-dimensional random short fiber composites requires only three independent variables to estimate Young's modulus ($E_{MML}$):

$$E_{MML} = \frac{1}{5} v_f E^f + (1 - v_f) E^m \qquad (106)$$

*Cox law*

This model [45] incorporates the effect of aspect ratio of reinforcement into Eq. (106). The following relation provides Young's modulus of the composites reinforced with random short fibers:

$$E_{COX} = \frac{1}{5}\left[1 - \frac{\tanh(\beta s)}{\beta s}\right] v_f E^f + (1 - v_f) E^m \qquad (107)$$

where $\quad s = \frac{4L_f}{d_f} \quad$ and $\quad \beta = \sqrt{\frac{2\pi E^m}{E^m(1 + v_m)\ln(1/v_f)}}$

*Randomly oriented chopped-fiber composites*

The basic constituents of SMC system are resin, filler, fiber and small amounts of additives. In determining the effective elastic properties of a typical SMC-composite, the direct contributions from the additives are ignored and their weights assigned to the resin weight [46]. By doing so, a filled SMC can be identified simply as a three-phase composite (fiber-filler-resin), and an unfilled SMC, a two-phase composite (fiber-resin).

When the resin and filler are mixed to form the matrix phase for the final composite, the effective bulk and shear moduli of the matrix (filled with filler and resin) can be obtained based on Hashin's expressions [47] as follows:

$$K^m = K^r + (K^p - K^r) \times \frac{(4G^r + 3K^r)v_p^*}{4G^r + 3K^p + 3(K^r - K^p)v_p^*} \tag{108}$$

$$G^m = G^r \left\{1 + \frac{15(1 - \nu^r)(G^p - G^r)G^r v_p^*}{(7 - 5\nu^r)G^r + 2(4 - 5\nu^r)\left[G^p - (G^p - G^r)v_p^*\right]}\right\} \tag{109}$$

with $v_p^*$ being the effective filler volume content of the matrix phase.

$$v_p^* = \frac{v_p}{v_p + v_r} \tag{110}$$

Young's modulus and Poisson's ratio, which are the effective stiffness properties needed to represent the filled matrix, can be derived as follows

$$E^m = \frac{9K^m G^m}{3K^m + G^m} \tag{111}$$

$$\nu^m = \frac{3K^m - 2G^m}{2(3K^m + G^m)} \tag{112}$$

Once the effective properties of the filled matrix phase are known, the three-phase composite can be represented as a fiber-matrix mixture. As derived by Hill [9, 15], axial Young's modulus and major Poisson's ratio for the uniaxial condition are given by:

$$E_1 = v_f E^f + v_m E^m + \frac{4 v_f v_m (\nu^f - \nu^m)^2}{\frac{\nu^m}{K^f + G^f/3} + \frac{\nu^f}{K^m + G^m/3} + \frac{1}{G^m}} \tag{113}$$

$$\nu_{12} = \nu^f v_f + \nu^m v_m + \left[v_f v_m (\nu^f - \nu^m) \left(\frac{\frac{1}{K^m + G^m/3} - \frac{1}{K^f + G^f/3}}{\frac{v_m}{K^f + G^f/3} + \frac{v_t}{K^m + G^m/3} + \frac{1}{G^m}}\right)\right] \tag{114}$$

where the superscript f denote the fiber, and the matrix volume fraction ($v_m$) is given by

$$v_m = v_r + v_p = 1 - v_f \tag{115}$$

Based on an elaborate variational principle developed by Hashin and Shtrikman [48], Hashin [49] obtained the following elastic coefficients.

$$G_{23} = G^m \left[ 1 + \left\{ \cfrac{v_f}{\cfrac{G^m}{G^f - G^m} + \cfrac{\left(K^m + \frac{7}{3}G^m\right)v_m}{2\left(K^m + \frac{4}{3}G^m\right)}} \right\} \right] \tag{116}$$

$$G_{12} = G^m \frac{G^f(1 + v_f) + G^m v_m}{G^f v_m + G^m(1 + v_f)} \tag{117}$$

$$K_{23} = K^m + \frac{G^m}{3} + \left\{ \cfrac{v_f}{\cfrac{1}{K^f - K^m + \frac{1}{3}(G^f - G^m)} + \cfrac{v_m}{K^m + \frac{4}{3}G^m}} \right\} \tag{118}$$

The crux of Christensen and Waal's [50] theory is that the stiffness of a two-dimensional, randomly oriented fiber composite can be generated from that of transversely isotropic, unidirectionaly aligned fiber composites by considering the fibers as evenly oriented from 0 to $\pi$. Adopting Cox's concept of integration, the elastic moduli of a two-dimensional composite are derived from those of a transversely isotropic, unidirectionaly aligned fiber composite as follows

$$P^c = \frac{1}{\pi} \int_0^\pi P(\theta) d\theta \tag{119}$$

where $P^c$ represents the elastic moduli of composite material and $P(\theta)$ represents the elastic moduli of an unidirectional composite material oriented at an angle $\theta$ with respect to the material axis.

Using Hill and Hashin's results given by Eqs. (113) and (114) and Eqs. (116) to (118), Christensen and Waals applied the concept of Eq. (119) to obtain the effective elastic properties of a two-dimensional randomly oriented fiber composite as follows

$$E^c = \frac{1}{u_1} - (u_1^2 - u_2^2) \tag{120}$$

$$\nu^c = \frac{u_2}{u_1} \tag{121}$$

where $E^c$ and $\nu^c$ denote Young's modulus and Poisson's ratio of the final composite.

$$u_1 = \frac{3}{8}E_1 + \frac{1}{2}G_{12} + \frac{(3 + 2\nu_{12} + \nu_{12}^2)G_{23}K_{23}}{2(G_{23} + K_{23})} \tag{122}$$

$$u_2 = \frac{1}{8}E_1 - \frac{1}{2}G_{12} + \frac{(1 + 6\nu_{12} + \nu_{12}^2)G_{23}K_{23}}{2(G_{23} + K_{23})} \tag{123}$$

The final composite density can be expressed by

$$\rho_c = (\rho_p - \rho_r)v_p^* v_m + \rho_r v_m + \rho_f v_f \tag{124}$$

The volume fractions of the constituents can be related to their given weight fractions as follow:

$$v_c = \frac{\rho_p \rho_f w_r}{\rho_p \rho_f w_r + \rho_f \rho_r w_p + \rho_r \rho_p w_r} \tag{125}$$

$$v_p = \frac{\rho_f \rho_r w_p}{\rho_p \rho_f w_r + \rho_f \rho_r w_p + \rho_r \rho_p w_f} \tag{126}$$

$$v_f = 1 - v_r - v_p \tag{127}$$

where $w_r$ represents the weight fraction of the $r^{th}$ phase.

*Strain concentration tensor for randomly dispersed fibers*

It may be noted that the elastic coefficient matrix [C] [see Eq. (63)] estimated by using two-phase MT method directly provides the effective elastic properties of the composite, where the fiber is aligned with the 1–axis. In case of random orientations of fibers, the terms enclosed with angle brackets in Eq. (64) represent the average value of the term over all orientations defined by transformation from the local coordinate system of the fibers to the global coordinate system. The transformed mechanical strain concentration tensor for the fibers with respect to the global coordinates is given by

$$[\tilde{A}_{ijkl}] = t_{ip}t_{jq}t_{kr}t_{ls}[A_{pqrs}] \tag{128}$$

where $t_{ij}$ are the direction cosines for the transformation and are given by

$$t_{11} = \cos\phi \cos\psi - \sin\phi \cos\gamma \sin\psi, \quad t_{12} = \sin\phi \cos\psi + \cos\phi \cos\gamma \sin\psi,$$

$$t_{13} = \sin\psi \sin\gamma, \quad t_{21} = -\cos\phi \sin\psi - \sin\phi \cos\gamma \cos\psi, \quad t_{22} = -\sin\phi \sin\psi + \cos\phi \cos\gamma \cos\psi,$$

$$t_{23} = \sin\gamma \cos\psi, \quad t_{31} = \sin\phi \sin\gamma, \quad t_{32} = -\cos\phi \sin\gamma \quad \text{and} \quad t_{33} = \cos\gamma$$

Consequently, the random orientation average of the dilute mechanical strain concentration tensor $\langle [A] \rangle$ can be determined by using the following equation [51]:

$$\langle [A] \rangle = \frac{\int_{-\pi}^{\pi} \int_{0}^{\pi} \int_{0}^{\pi/2} [\tilde{A}](\phi, \gamma, \psi) \sin\gamma \, d\phi d\gamma d\psi}{\int_{-\pi}^{\pi} \int_{0}^{\pi} \int_{0}^{\pi/2} \sin\gamma \, d\phi d\gamma d\psi} \tag{129}$$

where $\phi$, $\gamma$, and $\psi$ are the Euler angles with respect 1, 2 and 3 axes, respectively. It may be noted that the averaged mechanical strain concentration tensors given by Eqs. (64) and (129) are used for the cases of aligned and random orientations of fibers, respectively, in Eq. (63).

**Short fiber composite**

In the past, various micromechanics models, such as the dilute concentration model based on the Eshelby's equivalent inclusion, the self-consistent model for finite length fibers, MT models, bounding models, the Halpin-Tsai equations and shear lag model for estimating the effective properties of aligned short fiber composites were reviewed [52]. However, the MT model and Halpin-Tsai equations have been reported to be the efficient analytical approaches for predicting the effective properties of short fiber composites [52]. Earlier developed two- and three-phase MT models in previous sections can be used directly in property evaluation of short fiber composites. Therefore, only Halpin-Tsai equations are summarized in the following common form [53]:

$$\frac{P}{P^m} = \frac{1 + \xi \eta v_f}{1 - \eta v_f} \tag{130}$$

where

$$\eta = \frac{(P^f/P^m) - 1}{(P^f/P^m) + 1} \tag{131}$$

where P represents any one of the composite elastic moduli listed in Table 1; superscripts f and m refer to the fiber and matrix, respectively; and $\xi$ is a parameter that depends on the Poisson's ratio of matrix and on the particular elastic property being considered.

**Composite containing wavy fibers**

Fiber waviness in composite materials may occur as a result of a variety of manufacturing induced phenomena. For example, in filament winding, the winding pressure can influence the linearity of fibers in underlying layers [54]. The fiber waviness can be categorized as in-plane or out-of-plane. In-plane waviness involves the cooperative undulation of fibers in the plane of the lamina and out-of-plane waviness generally involves the cooperative undulation of multiple plies through the thickness of laminates [55]. Generally, in-plane waviness is found to be more severe than out-of-plane waviness [56]. When fiber waviness occurs, the mean fiber orientation usually remains parallel to the desired fiber direction, but displays some periodic curvature, frequently modeled as sinusoidal [57-66]. Therefore, the wavy fibers were modeled herein as sinusoidal fibers; while at any location along its length, the fiber is considered as transversely isotropic. It may be noted that the variations of the constructional feature of the composite can be such that the wavy fibers are coplanar with the 1–3 plane or the 2–3 plane as shown in Figs. 8 (a) and 8 (b), respectively. In the first, the amplitudes of the waves are parallel to the plane of a lamina while in the second, the amplitudes of the wavy fibers are normal to the plane of a lamina.

The RVE of the composite material containing a wavy fiber is illustrated in Fig. 9. As shown in this figure, the RVE is divided into infinitesimally thin slices of thickness dy. Averaging the effective properties of these slices over the length ($L_{RVE}$) of the RVE, the homogenized effective properties of the composite can be estimated. Each slice can be treated as an off-axis unidirectional lamina and its effective properties can be determined by transforming the effective properties of the corresponding orthotropic lamina. Now, these wavy fibers are characterized by

$$x = A\sin(\omega y) \quad \text{or} \quad z = A\sin(\omega y) \;;\quad \omega = n\pi/L_{RVE} \tag{132}$$

corresponding to the plane of fiber waviness being coplanar with the 1–3 plane and 2–3 plane, respectively. Here, A and $L_{RVE}$ denote the amplitude of the fiber wave and linear distance between the fiber ends, respectively; and n represents the number of waves of the fiber.

The running length ($l_{nr}$) of the fiber is given by:

$$L_{nr} = \int_0^{L_{RVE}} \sqrt{1 + A^2\omega^2\cos^2(\omega y)}\; dy \tag{133}$$

where the angle ɸ (shown in Fig. 9) is given by

$$\tan\phi = dx/dy = A\omega\cos(\omega y) \quad \text{or} \quad \tan\phi = dz/dy = A\omega\cos(\omega y) \quad (134)$$

corresponding to the plane of fiber waviness being coplanar with the 1–3 plane and 2–3 plane, respectively. Note that for a particular value of ω, the value of ϕ varies with the amplitude of fiber wave.

It may be noted that the effective thermoelastic properties at any point of any slice of the composite containing sinusoidally wavy fiber can be approximated by transforming the effective thermoelastic properties of the composite containing straight fibers. It may be noted that earlier obtained effective thermoelastic tensors ([C] and {α}; see Eqs. (51), (61), (63), (65), (68), (100) and (108)) are derived when the fibers are aligned along the 1-axis. These tensors should be transformed to obtain the properties when the fibers are aligned along the 3-axis using Eq. (102). Once $[C^{al}]$ and $\{\alpha^{al}\}$ are obtained, the effective elastic coefficient matrix $[C^C]$ and effective thermal expansion coefficient vector $\{\alpha^C\}$ at any point of any slice of the composite can be derived by employing the appropriate transformations. Transformations for the wavy fibers being coplanar with the plane of a composite lamina (i.e., 1-3 plane) are given by:

$$[C^C] = [T_1]^{-T}[C^{al}][T_1]^{-1} \quad \text{and} \quad \{\alpha^C\} = [T_1]^{-T}\{\alpha^{al}\} \quad (135)$$

in which

$$[T_1] = \begin{bmatrix} k^2 & 0 & l^2 & 0 & kl & 0 \\ 0 & 1 & 0 & 0 & 0 & 0 \\ l^2 & 0 & k^2 & 0 & -kl & 0 \\ 0 & 0 & 0 & k & 0 & -l \\ -2kl & 0 & 2kl & 0 & k^2-l^2 & -n \\ 0 & 0 & 0 & l & 0 & k \end{bmatrix}$$

$$k = \cos\phi = [1 + \{n\pi A/L_{RVE}\cos(n\pi y/L_{RVE})\}^2]^{-1/2} \quad \text{and}$$

$$l = \sin\phi = n\pi A/L_{RVE}\cos(n\pi y/L_{RVE})\,[1 + \{n\pi A/L_{RVE}\cos(n\pi y/L_{RVE})\}^2]^{-1/2}$$

Solving Eq. (135), we obtain the following relations

$$C_{11}^C = C_{11}^{al}k^4 + C_{33}^{al}l^4 + 2(C_{13}^{al} + 2C_{55}^{al})k^2l^2, \quad C_{12}^C = C_{12}^{al}k^2 + C_{23}^{al}l^2,$$

$$C_{13}^C = (C_{11}^{al} + C_{33}^{al} - 4C_{55}^{al})k^2l^2 + C_{13}^{al}(k^4 + l^4), \quad C_{22}^C = C_{22}^{al},$$

$$C_{23}^C = C_{12}^{al}l^2 + C_{23}^{al}k^2, \quad C_{33}^C = C_{11}^{al}l^4 + C_{33}^{al}k^4 + 2(C_{13}^{al} + 2C_{55}^{al})k^2l^2,$$

$$C_{44}^C = C_{44}^{al}k^2 + C_{66}^{al}l^2, \quad C_{55}^C = (C_{11}^{al} + C_{33}^{al} - 2C_{13}^{al} - 2C_{55}^{al})k^2l^2 + C_{55}^{al}(k^4 + l^4),$$

$$C_{66}^C = C_{44}^{al} l^2 + C_{66}^{al} k^2, \quad \alpha_{11}^C = \alpha_{11}^{al} k^2 + \alpha_{33}^{al} l^2, \quad \alpha_{22}^C = \alpha_{22}^{al} \quad \text{and} \quad \alpha_{33}^C = \alpha_{11}^{al} l^2 + \alpha_{33}^{al} k^2$$

The superscript al represents the properties of unidirectional composite lamina in which fibers are aligned along the 3-direction.

Similarly, if the plane of the fiber waviness is coplanar with the 2–3 plane, then the effective elastic ($C_{ij}^C$) and thermal expansion ($\alpha_{ij}^C$) coefficients at any point of the composite can be determined by using the following transformations:

$$[C^C] = [T_2]^{-T}[C^{al}][T_2]^{-1} \quad \text{and} \quad \{\alpha^C\} = [T_2]^{-T}\{\alpha^{al}\} \tag{136}$$

in which

$$[T_2] = \begin{bmatrix} 1 & 0 & 0 & 0 & 0 & 0 \\ 0 & k^2 & l^2 & kl & 0 & 0 \\ 0 & l^2 & k^2 & -kl & 0 & 0 \\ 0 & -2kl & 2kl & k^2-l^2 & 0 & 0 \\ 0 & 0 & 0 & 0 & k & -l \\ 0 & 0 & 0 & 0 & l & k \end{bmatrix}$$

Solving Eq. (136), we obtain the following relations

$$C_{11}^C = C_{11}^{al}, \quad C_{12}^C = C_{12}^{al} k^2 + C_{13}^{al} l^2, \quad C_{13}^C = C_{12}^{al} l^2 + C_{13}^{al} k^2,$$

$$C_{22}^C = C_{22}^{al} k^4 + C_{33}^{al} l^4 + 2(C_{23}^{al} + 2C_{44}^{al}) k^2 l^2,$$

$$C_{23}^C = (C_{22}^{al} + C_{33}^{al} - 4C_{44}^{al}) k^2 l^2 + C_{23}^{al}(k^4 + l^4),$$

$$C_{33}^C = C_{22}^{al} l^4 + C_{33}^{al} k^4 + 2(C_{23}^{al} + 2C_{44}^{al}) k^2 l^2,$$

$$C_{44}^C = (C_{22}^{al} + C_{33}^{al} - 2C_{23}^{al} - 2C_{44}^{al}) k^2 l^2 + C_{44}^{al}(k^4 + l^4),$$

$$C_{55}^C = C_{55}^{al} k^2 + C_{66}^{al} l^2, \quad C_{66}^C = C_{55}^{al} l^2 + C_{66}^{al} k^2,$$

$$\alpha_{11}^C = \alpha_{11}^{al}, \quad \alpha_{22}^C = \alpha_{22}^{al} k^2 + \alpha_{33}^{al} l^2 \quad \text{and} \quad \alpha_{33}^C = \alpha_{22}^{al} l^2 + \alpha_{33}^{al} k^2$$

It is now obvious that the effective thermoelastic properties of the composite lamina with the wavy fibers vary along the length of the fiber as the value of ϕ vary over the length of the fiber. The average effective elastic coefficient matrix $[\bar{C}^C]$ and thermal expansion coefficient vector $\{\bar{\alpha}^C\}$ of the lamina of such composite material can be obtained by averaging the transformed elastic $(C_{ij}^C)$ and thermal expansion $(\alpha_{ij}^C)$ coefficients over the linear distance between the fiber ends as follows [67]:

$$[\bar{C}^C] = \frac{1}{L_{RVE}} \int_0^{L_{RVE}} [C^C] \, dy \quad \text{and} \quad \{\bar{\alpha}^C\} = \frac{1}{L_{RVE}} \int_0^{L_{RVE}} \{\alpha^C\} \, dy \tag{137}$$

**The interphase model**

In order to determine the elastic properties of the interphase, several models have been implemented in earlier studies [68-75]. None of these models devoted attention to model the coefficients of thermal expansion of the interphase layer. The variation in the thermoelastic properties of the interphase is assumed to satisfy the following conditions,

$$[C^i]|_r = [C^f]|_{r=r_f}, \, [C^i]|_r = [C^m]|_{r=r_i}, \, \{\alpha^i\}|_r = \{\alpha^f\}|_{r=r_f} \text{ and } \{\alpha^i\}|_r = \{\alpha^m\}|_{r=r_i} \tag{138}$$

In Eq. (138), the respective superscripts f, i and m denote the fiber, interphase and matrix. Furthermore, $r_f$ is the fiber radius and $r_i$ is the outer radius of the interphase.

Recently, Kundalwal and Meguid [76] proposed a new interphase model to determine the thermoelastic coefficients of a nano-tailored composite satisfying the conditions given by Eq. (138). Our model would enable us to determine the thermoelastic properties of the fiber-matrix interphase; as follows,

$$[C^i]|_r = [C^m]\left(\frac{r_i}{r}\right) + \left[\left(\frac{r_i - r}{r_i - r_f}\right)\right]^\eta \left[[C^f] - [C^m]\left(\frac{r_i}{r_f}\right)\right] \tag{139}$$

$$\{\alpha^i\}|_r = \{\alpha^m\}\left(\frac{r_i}{r}\right) + \left[\left(\frac{r_i - r}{r_i - r_f}\right)\right]^\eta \left[\{\alpha^f\} - \{\alpha^m\}\left(\frac{r_i}{r_f}\right)\right] \tag{140}$$

where η is the adhesion exponent which controls the quality of adhesion between a fiber and the surrounding matrix.

Now, the effective thermoelastic properties of the interphase can be determined by averaging the varying interphase properties along radial direction; such that,

$$[C^i] = \frac{1}{r_i - r_f} \int_{r_f}^{r_i} [C^i]|_r \, dr \quad \text{and} \quad \{\alpha^i\} = \frac{1}{r_i - r_f} \int_{r_f}^{r_i} \{\alpha^i\}|_r \, dr \tag{141}$$

Note that the thermoelastic properties of the interphase were assumed to be vary along the radial direction using continuity conditions shown in Eq. (138). This assumption has been employed in several studies and validated with experimental and molecular dynamics results to estimate the interphase properties (see [76] and references therein).

**EFFECTIVE THERMAL CONDUCTIVITIES**

This section presents established micromechanics models for estimating the effective conductivities of a composite material containing either isotropic or orthotropic constituents.

**Effective medium approach (for isotropic constituents)**

First the Maxwell Garnett type effective medium approach was presented to estimate the effective thermal conductivities of the composite material incorporating the fiber-matrix interfacial thermal resistance. The effective medium approach by Nan et al. [77] can be modified to predict the effective thermal conductivities of the composite material reinforced with aligned fibers and are given by:

$$K_1 = v_f K^f + v_m K^m \tag{142}$$

$$K_2 = K_3 = K^m \frac{K^f(1+\alpha) + K^m + v_f[K^f(1-\alpha) - K^m]}{K^f(1+\alpha) + K^m - v_f[K^f(1-\alpha) - K^m]} \tag{143}$$

In Eq. (143), a dimensionless parameter $\alpha = 2a_k/d_f$ in which the interfacial thermal property is concentrated on a surface of zero thickness and characterized by Kaptiza radius, $a_k = R_k K^m$ where $d_f$ and $R_k$ represent the diameter of the fiber and fiber-matrix interfacial thermal resistance, respectively. The effective thermal conductivity matrix, [K], for the composite lamina can be represented as follows

$$[K] = \begin{bmatrix} K_1 & 0 & 0 \\ 0 & K_2 & 0 \\ 0 & 0 & K_3 \end{bmatrix} \tag{144}$$

The effective thermal conductivities ($\bar{K}_i$) at any point in the composite lamina where the fiber is inclined at an angle θ with the 1–axis (i.e., in 1–2 plane; see Fig. 7) can be derived in a straightforward manner by employing the appropriate transformation law as follows [78]:

$$[\bar{K}] = [T_1]^{-T}[K][T_1] \tag{145}$$

according as the fiber is coplanar with the 1–2 plane (see Fig. 1) and the corresponding transformation matrix is given by

$$[T_1] = \begin{bmatrix} \cos\theta & \sin\theta & 0 \\ -\sin\theta & \cos\theta & 0 \\ 0 & 0 & 1 \end{bmatrix}$$

**Composite cylinder assemblage approach (for isotropic constituents)**

This section presents the composite cylinder assemblage approach to estimate the effective thermal conductivities of the composite material. The effective thermal conductivities of the composite lamina are given by [79]:

$$K_1 = v_f K^f + (1 - v_f) K^m \tag{146}$$

$$K_2 = K_3 = K^m \left[ \frac{g(1 + v_f) + 1 - v_f}{g(1 - v_f) + 1 + v_f} \right] \tag{147}$$

where $g = K^f / K^m$

**Method of cells approach (for orthotropic constituents)**

The MOC approach by Aboudi et al. [80] can be modified to predict the effective thermal conductivities of the composite. Assuming that fibers as a rectangular cell, uniformly spaced in the matrix and are aligned along the $x_3$–axis, the composite can be viewed to be composed of cells forming doubly periodic arrays along the $x_1$– and $x_2$–directions. Figure 10 shows a repeating unit cell with four subcells.

Each rectangular subcell is labeled by β γ, with β and γ denoting the location of the subcell along the $x_1$– and $x_2$–directions, respectively. The subcell can be either a fiber or the matrix. Let four local coordinate systems ($\bar{x}_1^{(\beta)}, \bar{x}_2^{(\gamma)}$ and $x_3$) be introduced, all of which have origins that are located at the centroid of each cell. Here $b_\beta$, $h_\gamma$ and l denote the width, height and length of the subcell, respectively, while the volume (V) of the repeating unit cell is

$$V = bhl \tag{148}$$

Using MOC approach (for a detailed derivation of these equations the reader is referred to [80], the effective thermal conductivities of the composite lamina are given by

$$K_1^{al} = \frac{K_1^m\{K_1^f[h(V_{11} + V_{21}) + h_2(V_{12} + V_{22})] + K_1^m h_1(V_{12} + V_{22})\}}{hbl(K_1^m h_1 + K_1^f h_2)}$$

$$K_2^{al} = \frac{K_2^m\{K_2^f[b(V_{11} + V_{12}) + b_2(V_{21} + V_{22})] + K_2^m b_1(V_{21} + V_{22})\}}{hbl(K_2^m b_1 + K_2^f b_2)} \quad \text{and}$$

$$K_3^{al} = \frac{K_3^f V_{11} + K_3^m(V_{12} + (V_{21} + (V_{22}))}{hbl} \tag{149}$$

The effective thermal conductivities at any point in the composite where the fiber is inclined at an angle $\phi$ with the 3–axis (see Fig. 9) can be derived in a straightforward manner by employing the following appropriate transformation law as follows:

$$[K^C] = [T_1]^{-T}[K^{al}][T_1]^{-1} \quad \text{and} \quad [K^C] = [T_2]^{-T}[K^{al}][T_2]^{-1} \tag{150}$$

corresponding to the plane of fiber waviness being coplanar with the 1–3 plane and 2–3 plane, respectively (see Fig. 8). The various matrices appeared in Eq. (150) are given by

$$[K^{al}] = \begin{bmatrix} K_1^{al} & 0 & 0 \\ 0 & K_2^{al} & 0 \\ 0 & 0 & K_3^{al} \end{bmatrix}, \quad [T_1] = \begin{bmatrix} \cos\phi & 0 & \sin\phi \\ 0 & 0 & 0 \\ -\sin\phi & 0 & \cos\phi \end{bmatrix} \quad \text{and} \quad [T_2] = \begin{bmatrix} 1 & 0 & 0 \\ 0 & \cos\phi & \sin\phi \\ 0 & -\sin\phi & \cos\phi \end{bmatrix},$$

in which

$$k = \cos\phi = [1 + \{n\pi A/l_f \cos(n\pi y/l_f)\}^2]^{-1/2} \quad \text{and}$$

$$l = \sin\phi = n\pi A/l_f \cos(n\pi y/l_f)\,[1 + \{n\pi A/l_f \cos(n\pi y/l_f)\}^2]^{-1/2}$$

It is now obvious that the effective thermal conductivities of the composite lamina containing wavy fibers vary along the length of the fiber wave as the value of $\phi$ vary over its length. The average effective thermal conductivity matrix of the lamina of such composite containing wavy fibers can be obtained by averaging the transformed thermal conductivity coefficients over the linear distance between the fiber ends as follows:

$$[\bar{K}^C] = \frac{1}{L_{RVE}} \int_0^{L_{RVE}} [K^C]\,dy \tag{151}$$

## RESULTS AND DISCUSSIONS

In this section, numerical values of the effective properties of different composite systems are evaluated using the different models reviewed in the preceding sections. For clarity, only the established models were considered to predict the properties herein. The author also undertakes the some indicative comparisons with respective results obtained by other researchers. Finally, the application of micromechanical models has been presented to predict the properties of carbon nanotube reinforced composites.

## Effective thermoelastic properties

The natural questions to ask at this juncture are which model provides better estimates for thermoelastic properties of composites. Therefore, comparisons were made between the results predicted by the different models. The important associated aspects (fiber-matrix interphase, random and wavy orientations of fibers among others) of a composite in its property evaluation also considered while comparing the results.

### *Effect of interphase*

Let us first demonstrate the effect of fiber-matrix interphase on the effective thermoelastic properties of the silicon carbide fiber reinforced composite. For such investigation, the four models were considered; namely, two- and three-phase mechanics of materials (MOM) approaches, and two- and three-phase MT methods. Accordingly, the corresponding Eqs. (51), (61), (63) and (65) were used. Carbide fiber, fiber-matrix interphase and epoxy matrix are considered for evaluating the numerical results. Their material properties are listed in Table 2. Practically, the fiber volume fraction in the composites can vary typically from 0.3 to 0.7. Hence, this range was considered to analyze the effect of interphase on the effective thermoelastic coefficients. Using Eq. (141) and considering the respective values of adhesion coefficient ($\eta$) and thickness of interphase as 50 and 1 μm, the thermoelastic properties of fiber-matrix interphase (see Table 3) were determined. The lower value of $\eta$ indicates that a fiber is well bonded to the surrounding matrix through the intermediate interphase and the reverse is true for the higher value of $\eta$. Values of such adhesion coefficients can be determined experimentally using Kelly-Tyson model, Pukanszky model and pull-out model [31]. In several studies the effect of interphase on the nano- and micro-scale composites was studied, and readers are referred to Refs. [68-75, 81-84] for more details.

Figure 11 illustrates the variation of the effective elastic coefficient $C_{11}$ of the carbide fiber reinforced composite with the carbide fiber volume fraction ($v_f$). It can be observed that all the models predict identical estimates. It is important to note that the fiber-interphase does not affect the magnitude of $C_{11}$. Figure 12 illustrates the variation of the effective elastic coefficient $C_{12}$ of the composite with the values of $v_f$. It may be observed from Fig. 12 that (i) the MT method overestimates the values of $C_{12}$ in comparison to the MOM approach and (ii) the interphase slightly improves the values of $C_{12}$ for higher volume fraction of fiber. Since the composite is transversely isotropic, similar results are also obtained for the effective elastic coefficient $C_{13}$ but are not shown here. Similar trend of results can be observed for the effective transverse elastic coefficient $C_{22}$ of the composite as shown in Fig. 13. Since the composite is transversely isotropic with the 1– axis as the symmetry axis, the values of $C_{33}$ are found to be identical to those of $C_{22}$ and for brevity are not presented here. It may be observed from Fig. 14 that the three-phase MOM approach overestimates the values of $C_{23}$ in comparison to the two-phase MOM approach. On the other hand, MT method does not show the influence of interphase and it over predicts the results in comparison to the MOM approaches. The MT method significantly overestimates the values of $C_{55}$ of the composite as shown in Fig. 15; but the interphase does not influence this constant. Figure 16 illustrates the variation of the axial coefficient of thermal coefficient (CTE) $\alpha_1$ of the composite with the values of $v_f$. This figure reveals that the values of $\alpha_1$ decrease with the increase in the values of $v_f$; this is obvious result because the CTE of fiber is much lower than the other two constituents and therefore the values of $\alpha_1$ decrease as the fiber loading in the composite increases. It may also be observed from Fig. 16 that the interphase does not much influence the values of $\alpha_1$. Figure 17 demonstrate the linear decrease in the transverse CTE ($\alpha_2$) of the composite as the fiber loading in the composite increases. This figure also demonstrates that the three-phase MOM approach slightly over predicts the values of $\alpha_2$ in comparison to all other models. It may be observed from Figs. 11-17 that the incorporation of an interphase between the fiber and matrix slightly increases the transverse thermoelastic coefficients of the resulting composite; this finding is coherent with the previously reported results [76, 87-91]. Comparison of the results predicted by MOM approach with that by the MT methods reveals that the MOM approach yields conservative estimates;

hence, predictions by the MOM approach have been considered in the subsequent results for investigating the effect of interphase thickness and adhesion coefficient on the effective thermoelastic properties of the composite.

Practically, the fiber-matrix interphase thickness in the composite can vary; therefore, the investigation of the effect of interphase thickness on the effective thermoelastic properties of the composite would be an important study. For such investigation, two discrete values of interphase thicknesses (t) and adhesion coefficients ($\eta$) are considered: $t = 1, 5$ µm and $\eta = 5, 10$. Figures 18-23 illustrate the comparisons of the thermoelastic constants of the composite for different values of t and $\eta$. It may be observed from these figures that the values of thermoelastic coefficients, except the axial elastic coefficient ($C_{11}$), are slightly improved with the increase in the interphase thickness at higher loading of fibres.

*Effect of random orientation of fibers*

Practically, the orientations of the fiber reinforcement in the matrix can vary over the volume of the composite. Therefore, studying the properties of composites reinforced with randomly oriented fibers is of a great importance. For such investigation, carbide fibers are considered to be randomly dispersed in the matrix over the volume of the composite. First the effective thermoelastic properties of the unidirectional fiber reinforced composite were determined using the two-phase MT method; subsequently, the transformed mechanical strain concentration tensor given by Eq. (129) is used to estimate the average properties of composite containing randomly dispersed fibers. As expected, this case provides the isotropic thermoelastic properties for the resulting composite. Figures 24-26 illustrate the variations of the effective thermoelastic coefficients of the composite with the fiber loading. These results clearly demonstrate that the randomly dispersed fibers improve the values of $C_{12}$, $C_{22}$ and $\alpha_1$ of the composite over those of the values of $C_{12}$, $C_{22}$ and $\alpha_1$ of the composite reinforced with aligned fibers (see Figs. 12 and 13); the respective improvements are found to be 470%, 600% and 74% when the fiber loading is 0.7. On the other hand, it may be observed that the random orientations of fibers significantly affect the effective CTEs of the composite. The improvement in the elastic properties is attributed to the fact that the fibers are homogeneously dispersed in the matrix in the random case and hence the overall elastic properties of the resulting composite improve in comparison to the aligned case.

*Effect of waviness of fibers*

The effect of fiber waviness on the effective thermoelastic properties of the composite is investigated when wavy fibers are coplanar with either of the two mutually orthogonal planes. It may be noted that the axis symmetry of carbide fibers is considered to be aligned along the 3–direction (see Figs. 8 and 9); accordingly, the transformations for the wavy fibers being coplanar with the 1-3 plane or the 2-3 plane are carried out. The effect of fiber waviness is studied by varying the value of fiber wave frequency, $\omega = n\pi/L_n$; the multiplying factor (n) takes the value from 0 to 24 and the amplitude (A) of the fiber wave is considered as 75 μm. Figure 27 illustrates the variations of the values of $C_{11}$ of the composite with the wave frequency. It may be observed from Fig. 27 that the value of $C_{11}$ significantly increases as the value of ω increases when the wavy fibers are coplanar with the 1–3 plane but the values of $C_{11}$ are not much influenced by the wavy fibers being coplanar in the 2–3 plane. Although not shown here, the same trend of results are obtained for the values of $C_{12}$. On the other hand, it may be observed from Fig. 28 that the values of $C_{22}$ significantly increase as the value of ω increases when the wavy fibers are coplanar with the 2–3 plane but the values of $C_{22}$ are not much influenced by the wavy fibers being coplanar in the 1–3 plane. Similar trend of results has been obtained for the values of $C_{23}$ as shown in Fig. 29. Figure 30 demonstrates that the value of $C_{55}$ significantly enhances when the wavy fibers are coplanar with the 1–3 plane. Since the composite is transversely isotropic material, the values of $C_{66}$ are found to be identical to those of the values of $C_{55}$. Figures 31 and 32 demonstrate the variations of the values of $\alpha_1$ and $\alpha_2$ of the composite with the wave frequency. These figures reveal that the values of $\alpha_1$ and $\alpha_2$ increase with the increase in the values of ω when the wavy fibers are coplanar with the 1–3 plane and 2–3 plane, respectively. It may be noted from Figs. 27–32 that if the wavy fibers are coplanar with the 1–3 plane then the axial thermoelastic coefficients of the composite are significantly improved. When the wavy fibers are coplanar with the plane 1–3 as shown in Fig. 8(a), the amplitudes of the fiber waves becomes parallel to the 1–axis and this results into the aligning of the projections of parts of fibers lengths with the 1–axis leading to the axial stiffening of the composite. The more is the value of ω, the more will be such projections and hence the effective axial thermoelastic coefficients ($C_{11}$, $C_{12}$, $C_{13}$, $C_{55}$, $C_{66}$, and $\alpha_1$) of the composite increase with the increase in the values of ω. On the other hand, if the wavy fibers are coplanar with the transverse plane (2–3 plane) then the transverse thermoelastic coefficients ($C_{22}$, $C_{23}$, $C_{33}$, $C_{44}$, $\alpha_2$, and $\alpha_3$) of the composite are improved.

**Application of micromechanics models to nanocomposites**

Polymer nanocomposites embedded with nanoparticles such as carbon nanotubes (CNTs), graphene and their derivatives, nanoclays, and silica nanoparticles have attracted a large amount of attention to achieve more improved thermomechanical properties than conventional composites [92-105]. Among all others, CNTs [106] have been emerged as the ideal candidates for multifarious applications due to their remarkable thermoelastic and physical properties. The quest for utilizing the remarkable thermomechanical properties of CNTs [90, 107-109] has led to the emergence of a new area of research that involves the development of nanocomposites [110-122]. These studies revealed that the addition of a small fraction of CNTs into a polymer matrix introduces significant improvement in the multifunctional properties of micro- and nano-nanocomposites. Therefore, the current study endeavoured to utilize the micromechanics models for estimating the thermoelastic properties of CNT-reinforced nanocomposites herein.

*Thermoelastic properties*

The performance of a CNT-reinforced nanocomposite material is critically controlled by the interfacial bonding between a CNT and the surrounding matrix material. Chang et al. [123] used Raman scattering and X-ray diffraction techniques, and showed that no chemical bonding exists between a CNT and the surrounding polymer matrix. In this case, only the dominating non-bonded van der Waals (vdW) interactions can be considered between a CNT and the surrounding polymer matrix [124-126]. In order to estimate the effective thermoelastic properties of the CNT-reinforced composite, the consideration of vdW interactions between a CNT and the surrounding polymer matrix, is an important issue. In several research studies [90, 110, 120], an equivalent solid continuum interphase was considered between a CNT and the polymer matrix which characterizes vdW interactions; this assumption allows us to use the interphase model utilized in the current study for estimating the thermoelastic properties of the CNT-matrix interphase. The results predicted by the interphase model were compared with the predictions of molecular dynamics simulations carried out by Tsai et al. [90]. In their study, the thickness of the interphase ($h_i$) was assumed equal to the non-bonded gap, and the corresponding elastic properties were determined from the molecular energy calculated from molecular dynamics simulations. Tsai et al. [90] determined Young's modulus of the interphase from the normalized non-bonded energy in CNT-polyimide nanocomposite considering three different types of zigzag CNTs. For the comparison purpose, transversely isotropic elastic properties of zigzag CNTs are taken from Tsai et al. [90] and the elastic properties of the polymer material are taken from Odegard et al. [110]. In Table 4, the predicted Young's modulus of the interphase considering adhesion coefficient ($\eta$) as 40 are compared with the existing molecular dynamics simulations.

With CNT type varying from (10, 0) to (18, 0), these comparisons reveal that: (i) the interphase model accurately predicts Young's modulus of the interphase for the CNT (14, 0) and (ii) over predicts Young's modulus of the interphase for a smaller diameter of CNT and (iii) under predicts the same for larger diameter of CNT. The marginal differences observed are attributed to the fact that the polyimide polymer is generated by 10 repeated monomer units in the study of Tsai et al. [90], whereas in the interphase model, the elastic properties of the polyimide material are taken from Odegard et al. [110]. Thus, it can be inferred from these comparisons that the interphase model can be reliably applied to predict the thermoelastic properties of the interphase if the value of CNT diameter is between 1 - 1.4 nm and the value of $\eta$ is 40.

It is accepted by many researchers that CNT-reinforced composites can be considered as fiber reinforced composites so that their elastic properties can be predicted by using the available micromechanics methods [120]. Seidel and Lagoudas [87] estimated the effective elastic properties of CNT-reinforced composites employing the self-consistent and the MT methods. Song and Youn [127] numerically estimated the effective elastic properties of CNT-reinforced polymer based composites by using the homogenization technique. The control volume finite element method is adopted in their study to implement the homogenization method with the assumption that the CNT-epoxy nanocomposites have geometrical periodicity with respect to a microscopic scale. Recently, Farsadi et al. [128] developed a three-dimensional finite element model to investigate the effect of CNT volume fraction and waviness on the mechanical properties of nanocomposites reinforced with waved CNT. They reported that the CNT waviness decreases the axial and transverse Young's moduli of nanocomposites but the change in the value of transverse Young's modulus is less remarkable than the axial Young's modulus.

Suggestive literature indicates that the established micromechanics models can be utilized to determine the effective properties of CNT-reinforced nanocomposites. Therefore, the MT method was utilized herein to determine the effective thermoelastic properties of CNT-reinforced nanocomposites and the obtained results were compared with those of the obtained by Farsadi et al. [128]. Table 5 summarises the outcome of this comparison. The two sets of results predicted by the finite element and the MT model are in excellent agreement, validating the later.

To further show the applicability of the MT method, the effective CTEs of the CNT-reinforced nanocomposite predicted by the MT method are compared with the finite element predictions of the nanocomposite containing straight CNTs. Kirtania and Chakraborty [129] determined the CTEs of the aligned CNT-reinforced nanocomposite by employing the finite element model. For the comparison purpose, CNTs and matrix material of the nanocomposite studied by Kirtania and Chakraborty [129] are considered for the constituent phases of the nanocomposite. For the effective CTEs of the CNT-reinforced nanocomposite, the two sets of results are in excellent agreement as demonstrated in Table 6.

*Thermal conductivities*

In this section, the effective thermal conductivities of CNT-reinforced nanocomposite are presented using the effective medium (EM) and the composite cylinder assemblage (CCA) approaches. The current predictions are also compared against those from the experimental results by Marconnet et al. [130]. They fabricated the aligned CNT-polymer nanocomposites consisting of CNT arrays infiltrated with an aerospace-grade thermoset epoxy. For the comparison purpose, the values of thermal conductivities of the CNT and polymer matrix are taken as $K^n = 22.1$ W/mK and $K^p = 0.26$ W/mK, respectively, as considered by Marconnet et al. [130]. The comparisons of the axial ($K_A$) and transverse ($K_T$) thermal conductivities of the aligned CNT-polymer nanocomposites estimated by the EM and CCA approaches with those of the experimental results are illustrated in Figs. 33 and 34, respectively. In these figures, dotted blue line represents best fits obtained from the EM approach for the experimental results considering an alignment factor (AF) of CNTs as 0.77 observed from the scanning electron microscopy (SEM). Figure 33 reveal that the values of $K_A$ predicted by the EM and CCA approaches overestimate the values of $K_A$ by 20% and 25% when the values of CNT volume fractions are 0.07 and 0.16, respectively. On the other hand, the EM and CCA approaches underestimate the values of $K_T$ by 40% and 58% when the values of CNT volume fractions are 0.07 and 0.16, respectively. These differences between the results are attributed to the fact that the perfect alignments of CNTs (i.e., AF = 1) are considered while computing the results by the EM and CCA approaches whereas the value of AF is 0.77 in Ref. [130]. Other possible reasons for the disparity between the analytical and experimental results include the CNT-matrix interfacial thermal resistance [130, 131], lattice defects within CNTs [132], and modification of the phonon conduction within CNTs due to interactions with the matrix [133]. It can be inferred from the comparisons shown in Figs. 33 and 34 that the EM and CCA approaches can be reasonably applied to predict the thermal conductivities of the CNT-reinforced nanocomposites.

*Advanced nanocomposites*

Literature on hybrid composites ([120] and references therein) indicate that the use of CNTs and conventional fibers together, as multiscale reinforcements, significantly improves the overall properties of resulting hybrid composites, which are unseen in conventional composites. More success can be achieved in improving the multifunctional properties of the hybrid composites by growing CNTs on the surfaces of conventional fibers. For instance, a novel nano-reinforced laminated composite (NRLC), containing carbon fibers grown with the CNTs, has been characterized experimentally and numerically by Kulkarni et al. [111]. The longitudinal and transverse cross sections of such NRLC are shown schematically in Fig. 35.

Therefore, to show the applicability of micromechanics models to advanced composites, the results predicted by the MT method are compared with those of numerical and experimental findings on the NRLC. The constituent phases as well as the geometrical parameters of the NRLC studied by Kulkarni et al. [111] are considered for the comparison purpose. Table 7 demonstrates the comparison between the two sets of results for the NRLC. It may be observed from Table 7 that the predicted value of the transverse Young's modulus ($E_x$) of the NRLC computed by the MT method match closely with that of the experimental value. The experimental value of $E_x$ is lower than the theoretical prediction, and this is attributed to the fact that CNTs are not perfectly radially grown and straight, and hence the radial stiffening of the NRLC decreases [111]. It may also be noted that the value of $E_x$ predicted by the MT method is much closer to the experimental value than that of the numerical value predicted by Kulkarni et al. [111]. These comparisons are significant since the prediction of the transverse Young's modulus of the NRLC provides critical check on the validity of the MT method. Thus, it can be inferred from the comparisons shown in Table 7 that the micromechanics models can be reliably applied to predict the overall thermoelastic properties of the advanced composite.

## SUMMARY

This article reviews and evaluates several micromechanics models that predict the thermomechanical properties of nano- and micro-fiber reinforced composite materials. The overall goal of this work is to provide quick and intuitive property predictions for different composite systems using established analytical micromechanical models. In order to investigate the effect of reinforcement on the thermomechanical properties, different cases of reinforcements are considered: aligned, random, and wavy. Specific attention was given to investigate the effect of fiber waviness on the effective thermoelastic properties of composite when wavy fibers are coplanar with either of the two mutually orthogonal planes. The current study shows the application of interphase model, which is developed by the current author in his previous work, to predict the thermoelastic properties of fiber-matrix interphase. This work also shows the applicability of established micromechanics models to predict the properties of CNT-reinforced composites considering the non-bonded interactions between a CNT and the polymer matrix as an equivalent solid continuum. The results from the micromechanics models are validated with the existing numerical and experimental findings on the nanocomposites and found to be in good agreement. The following is a summary of findings of the current study:
1. Mechanics of materials approach yields conservative estimates compared to the Mori-Tanaka approach,
2. Pronounced effect of fiber-matrix interphase on the effective transverse thermoelastic coefficients of the fiber reinforced composite is observed at higher volume fraction of fibers,
3. Randomly oriented fibers have significant influence on the thermoelastic properties of the composite,
4. The transverse thermoelastic coefficients of composites reinforced with aligned fibers are less than those of the randomly dispersed case,
5. Significant improvement has been observed in the effective thermoelastic properties of the composite when wavy fibers are coplanar with either of the two mutually orthogonal planes;

the axial and transverse thermoelastic properties of a composite improve when wavy fibers are coplanar with the plane of a lamina and the transverse plane of a lamina, respectively, and

6. The modeling strategy developed herein is capable of determining the effective thermoelastic properties and thermal conductivities of any advanced composite containing aligned or randomly dispersed nano- and micro-fibers with less computational efforts.

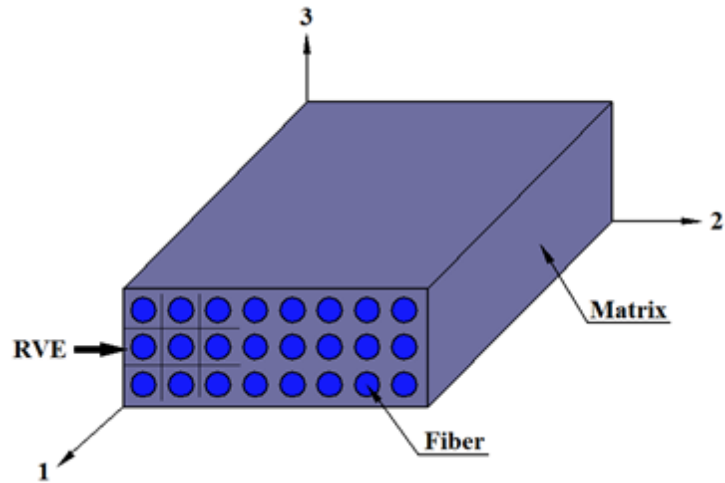

FIG. 1. A lamina of composite material illustrating representative volume element (RVE).

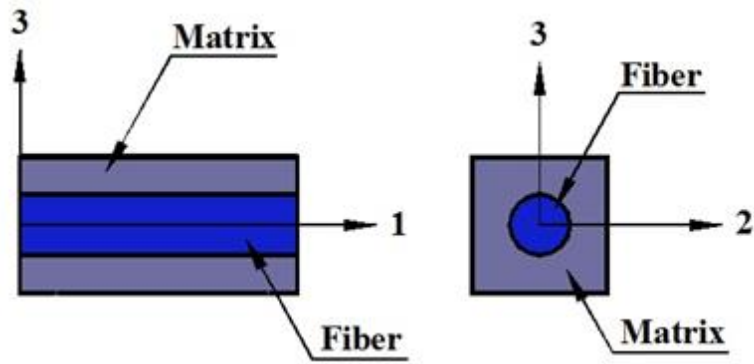

FIG. 2. Cross sections of the RVE of two-phase composite material.

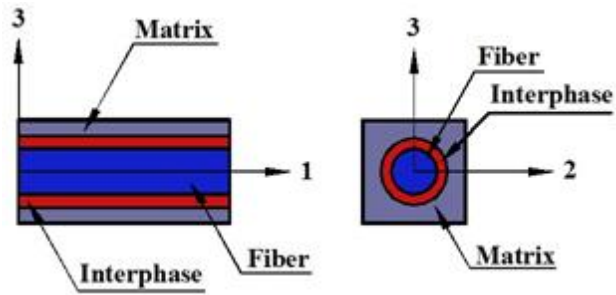

FIG. 3. Cross sections of the RVE of three-phase composite material.

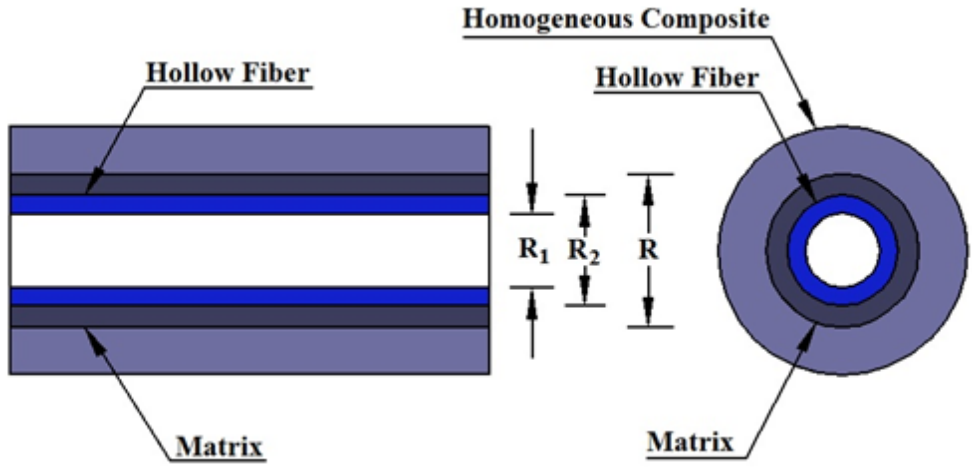

FIG. 4. RVE of the composite cylindrical assemblage model.

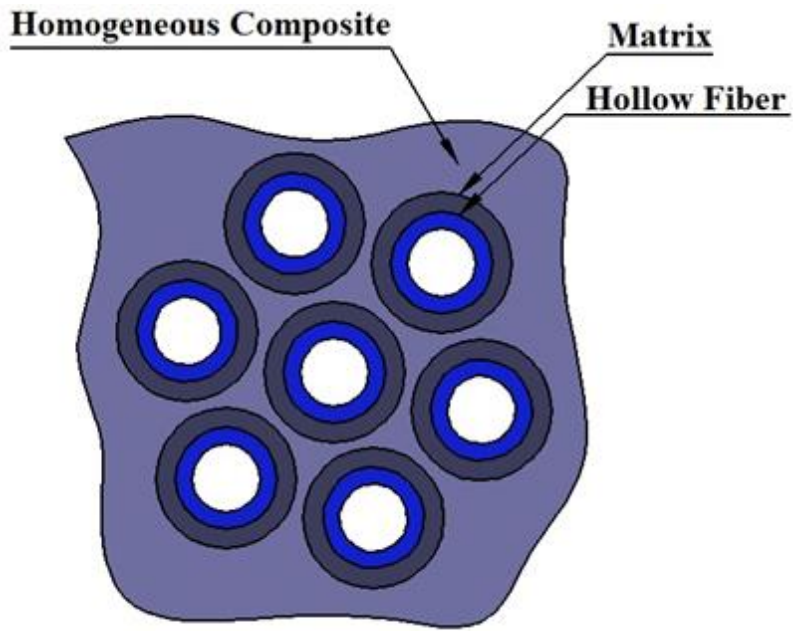

FIG. 5. Composite cylindrical assemblage made of hexagonal array of hollow circular fibers.

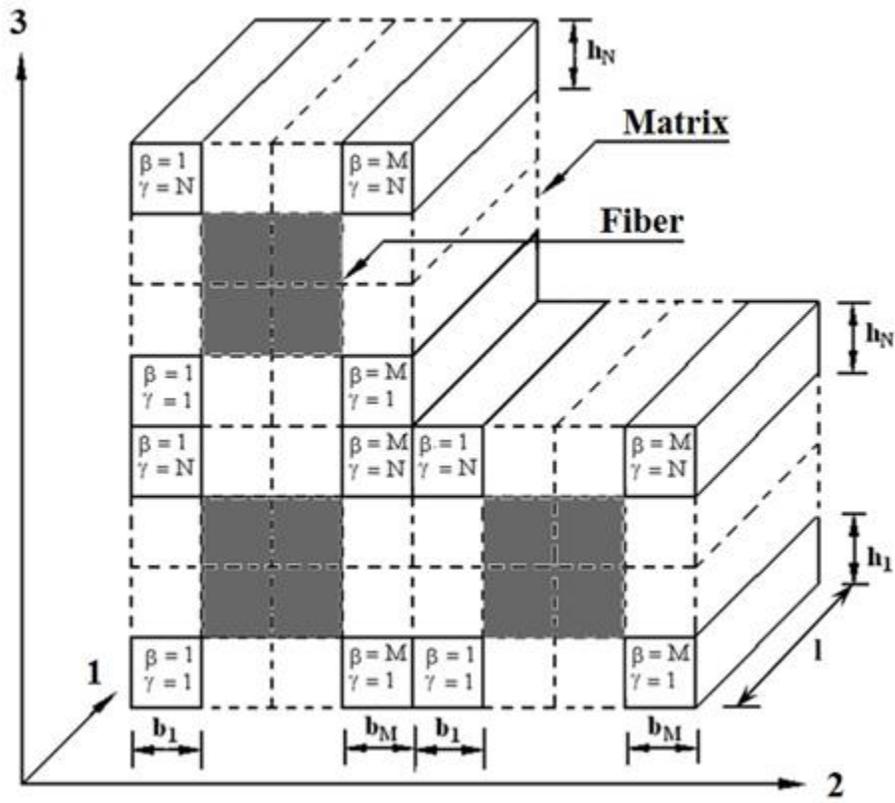

FIG. 6. Representative unit cell of the composite.

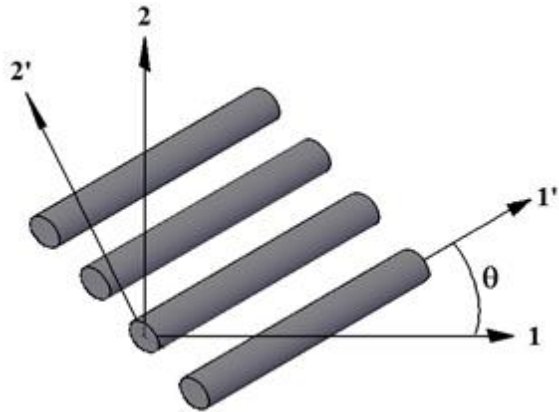

FIG. 7. Rotation of principal axes from 1-2 axes.

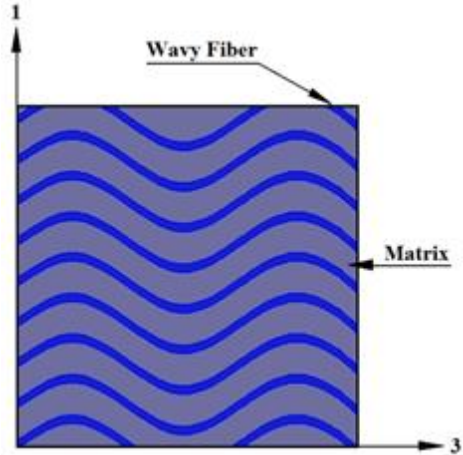

FIG. 8A. Composite material containing a wavy fiber coplanar with the plane of a lamina (i.e., 1–3 plane).

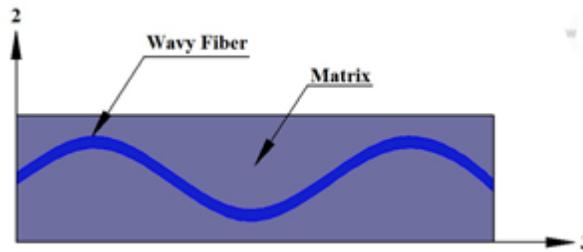

FIG. 8B. Composite material containing a wavy fiber coplanar with the transverse plane of a lamina (i.e., 2–3 plane).

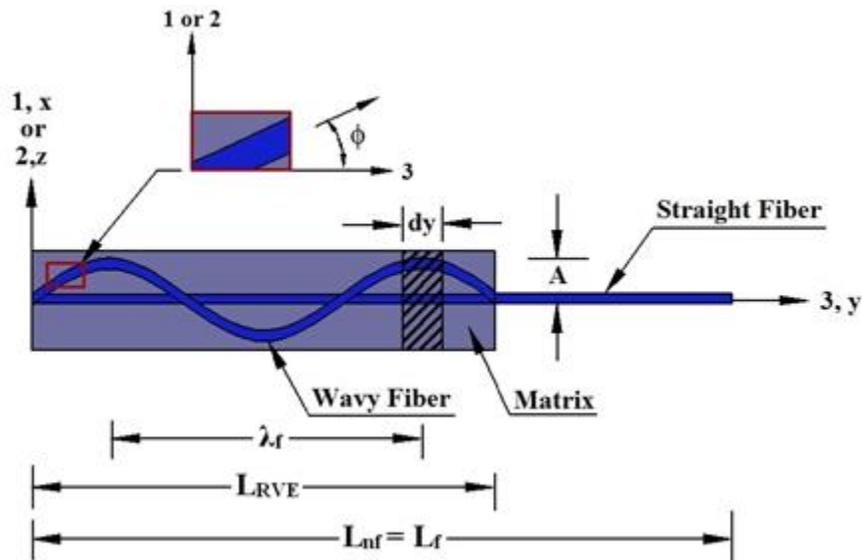

FIG. 9. RVE of the composite material containing a wavy fiber coplanar with either the plane of a lamina (i.e., 1–3 plane) or the transverse plane of a lamina (i.e., 2–3 plane).

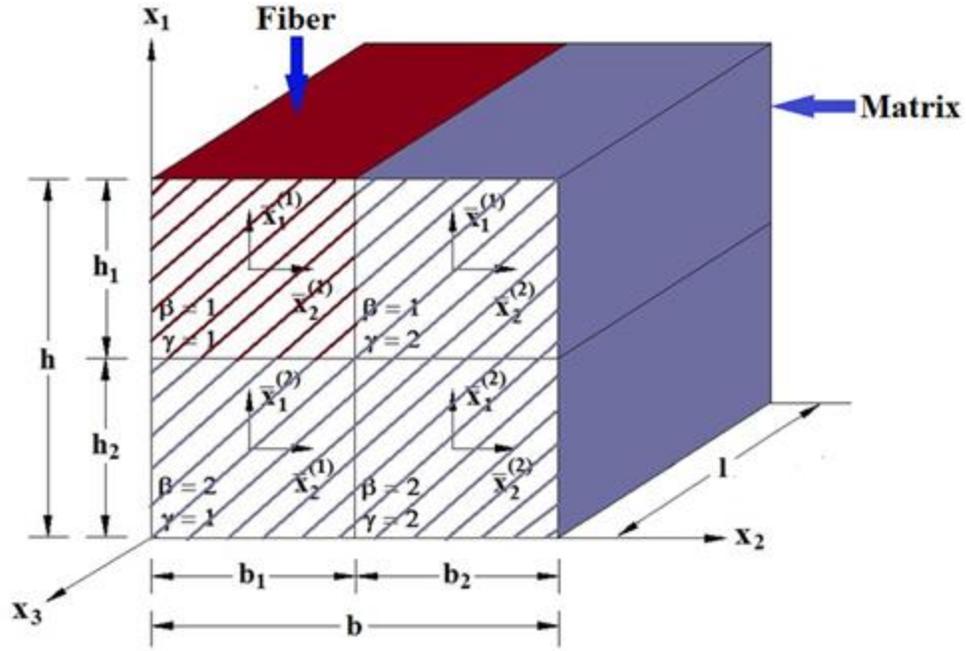

FIG. 10. Repeating unit cell of the composite material with four subcells ($\beta, \gamma = 1, 2$).

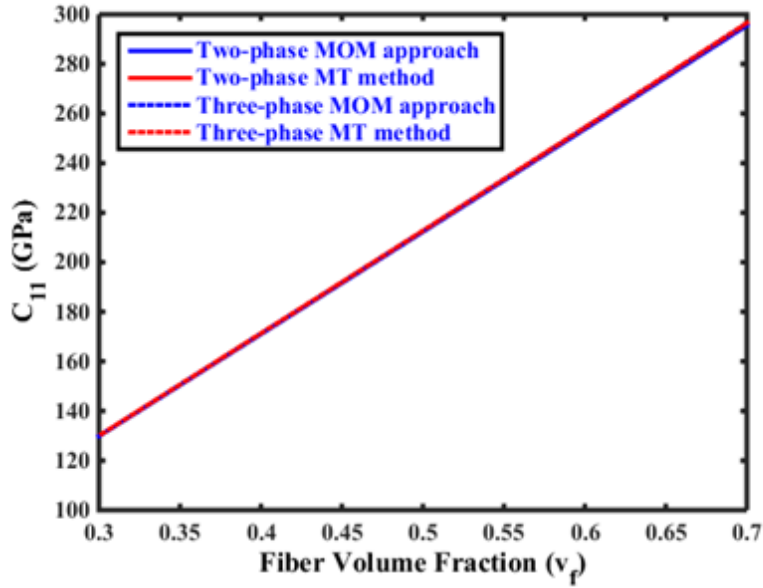

FIG. 11. Variation of the effective elastic coefficient $C_{11}$ of the composite against carbide fiber volume fraction.

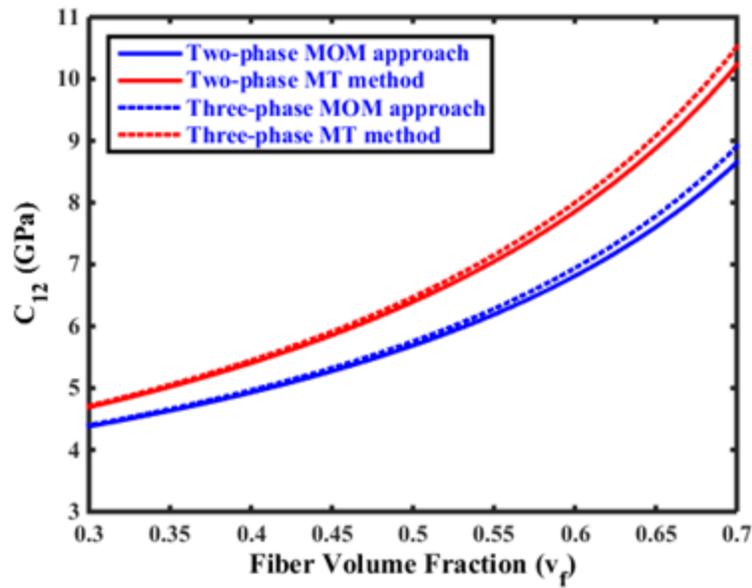

FIG. 12. Variation of the effective elastic coefficient $C_{12}$ of the composite against carbide fiber volume fraction.

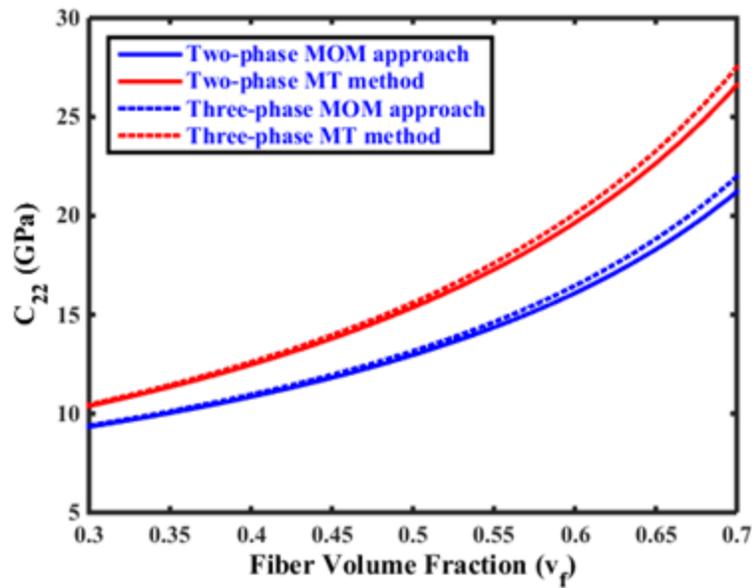

FIG. 13. Variation of the effective elastic coefficient $C_{22}$ of the composite against carbide fiber volume fraction.

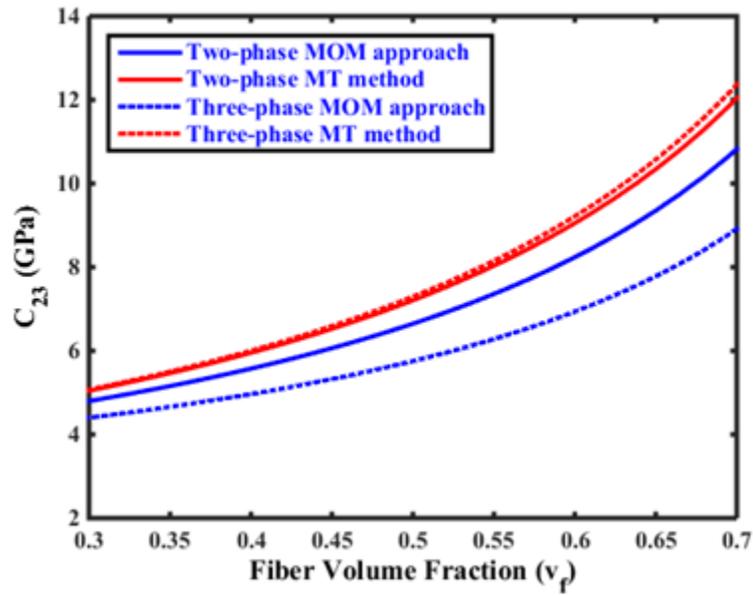

FIG. 14. Variation of the effective elastic coefficient $C_{23}$ of the composite against carbide fiber volume fraction.

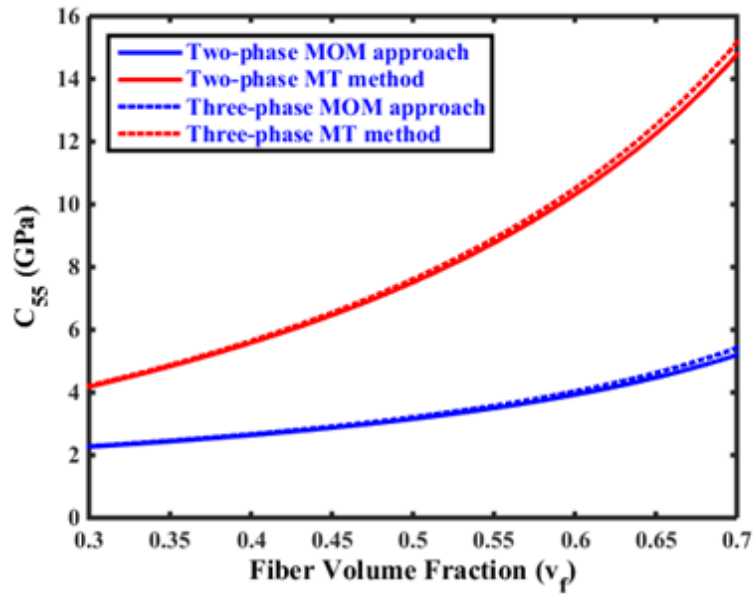

FIG. 15. Variation of the effective elastic coefficient $C_{55}$ of the composite against carbide fiber volume fraction.

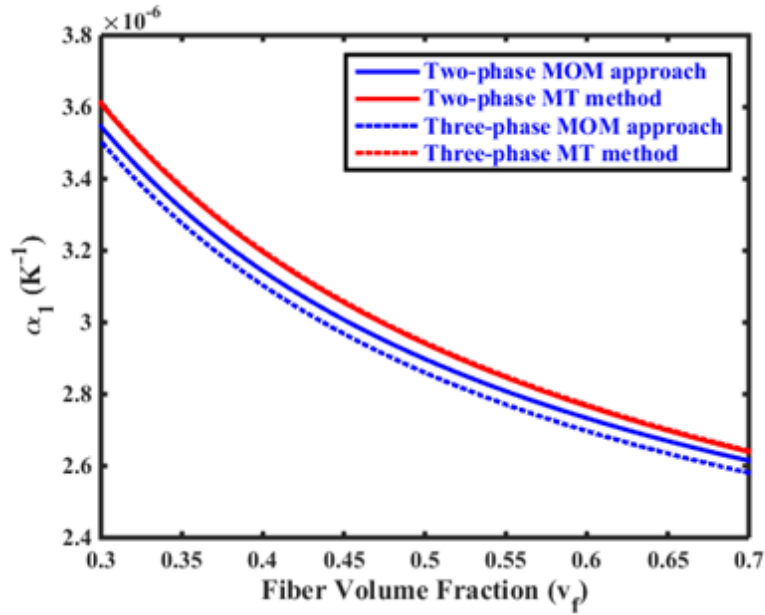

FIG. 16. Variation of the axial CTE $\alpha_1$ of the composite against carbide fiber volume fraction.

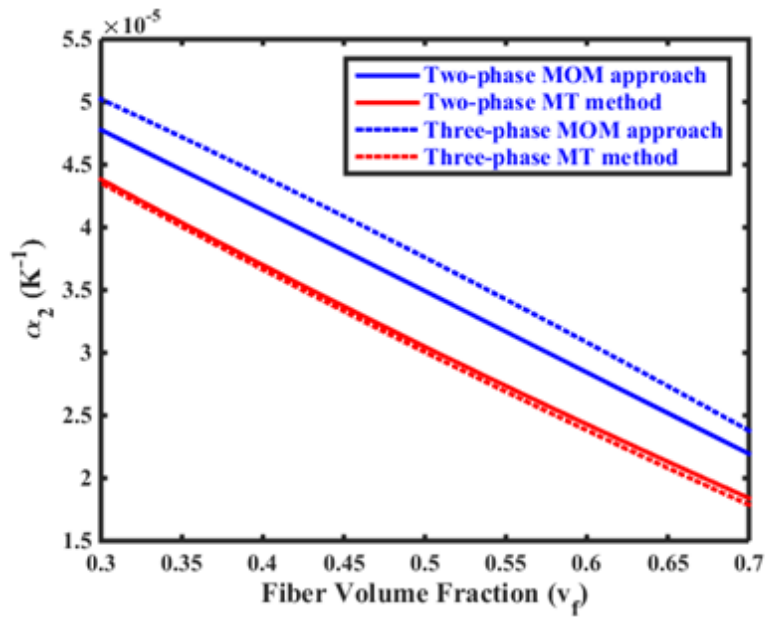

FIG. 17. Variation of the transverse CTE $\alpha_2$ of the composite against carbide fiber volume fraction.

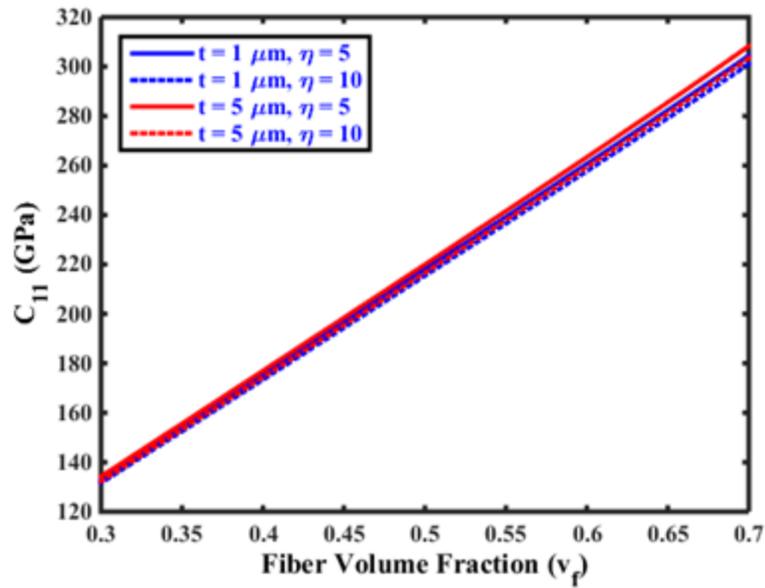

FIG. 18. Effect of interphase thickness on the effective elastic coefficient $C_{11}$ of the three-phase composite.

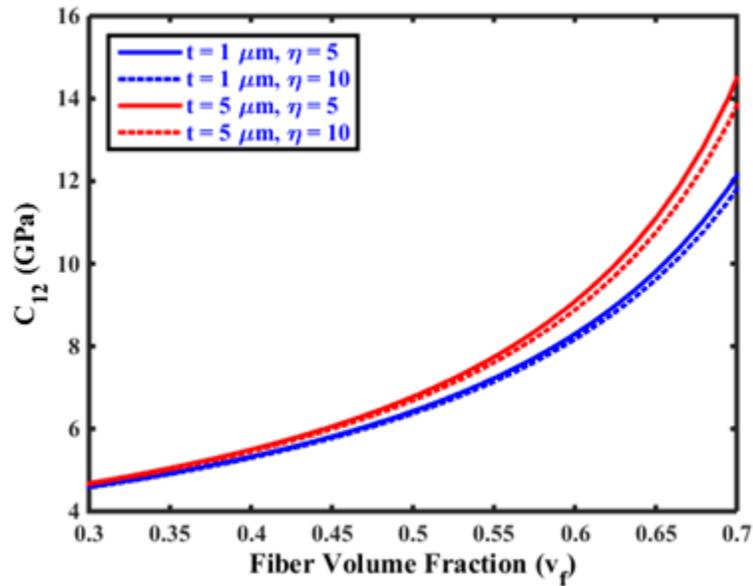

FIG. 19. Effect of interphase thickness on the effective elastic coefficient $C_{12}$ of the three-phase composite.

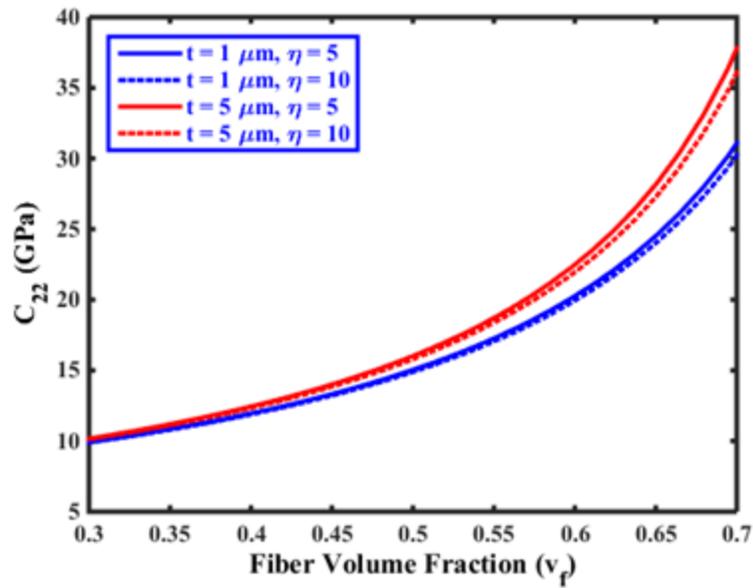

FIG. 20. Effect of interphase thickness on the effective elastic coefficient $C_{22}$ of the three-phase composite.

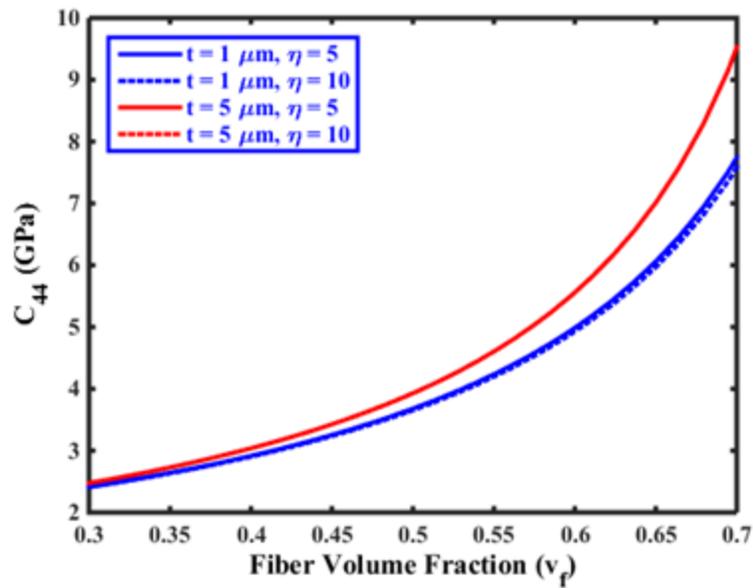

FIG. 21. Effect of interphase thickness on the effective elastic coefficient $C_{44}$ of the three-phase composite.

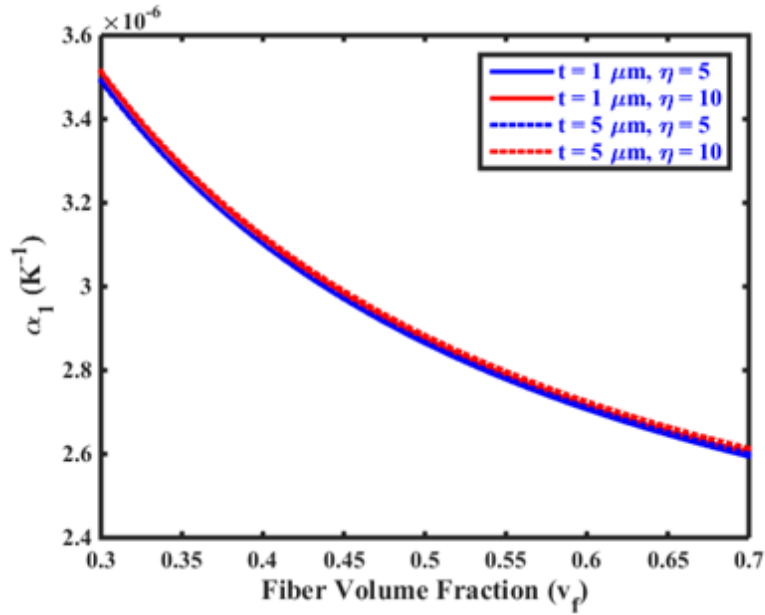

FIG. 22. Effect of interphase thickness on the effective axial CTE $\alpha_1$ of the three-phase composite.

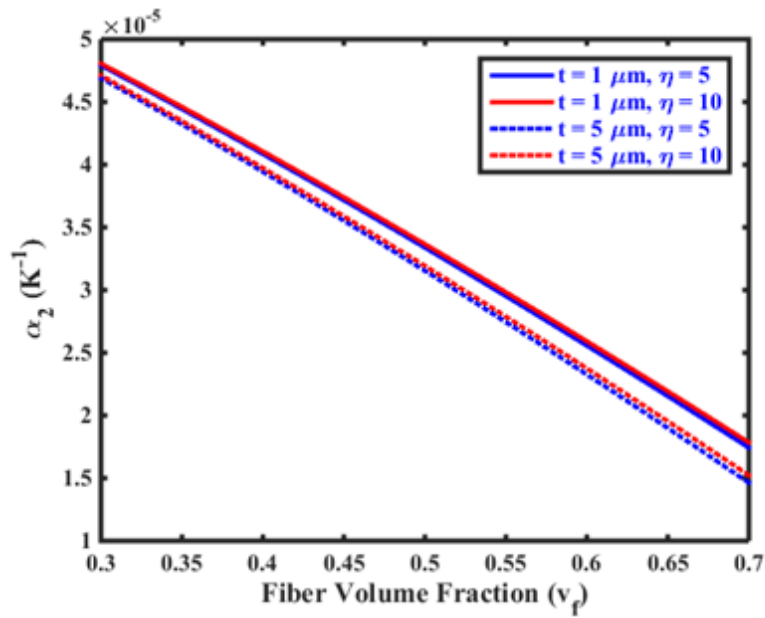

FIG. 23. Effect of interphase thickness on the effective transverse CTE $\alpha_2$ of the three-phase composite.

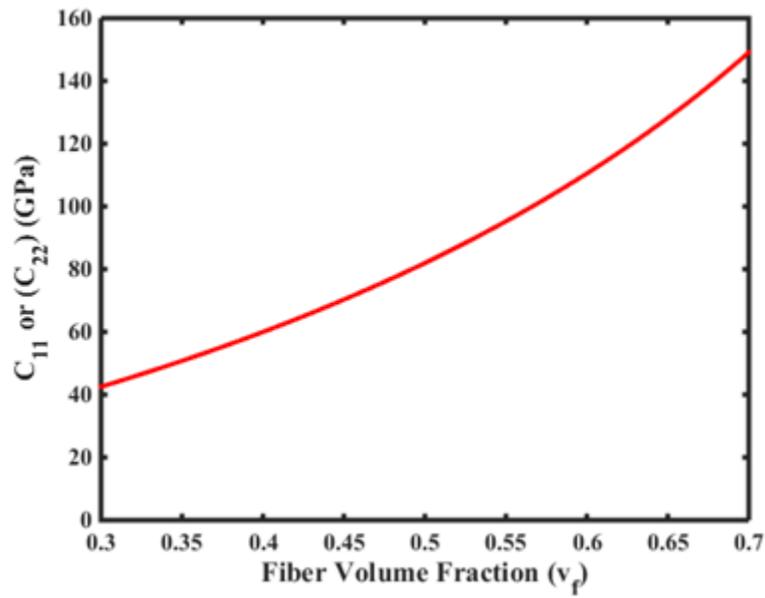

FIG. 24. Variation of the effective elastic coefficient $C_{11}$ (or $C_{22}$) of the composite containing randomly dispersed fibers.

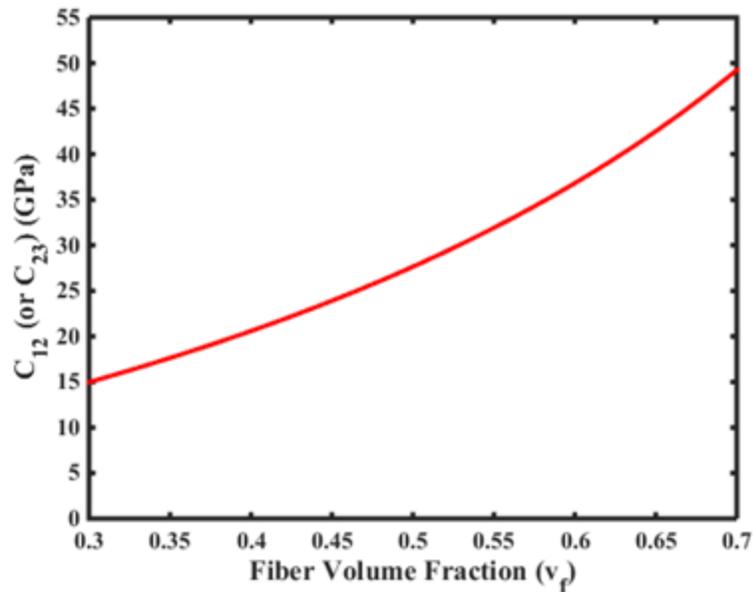

FIG. 25. Variation of the effective elastic coefficient $C_{12}$ (or $C_{23}$) of the composite containing randomly dispersed fibers.

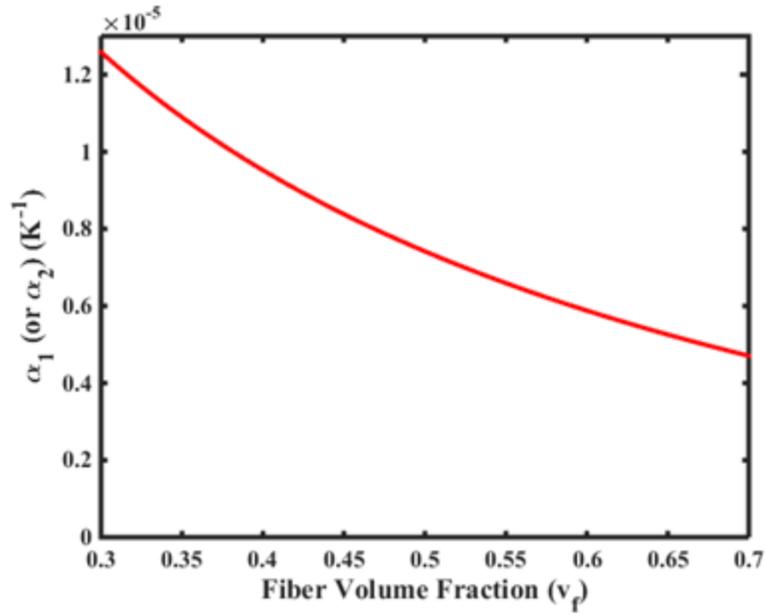

FIG. 26. Variation of the effective CTE $\alpha_1$ (or $\alpha_2$) of the composite containing randomly dispersed fibers.

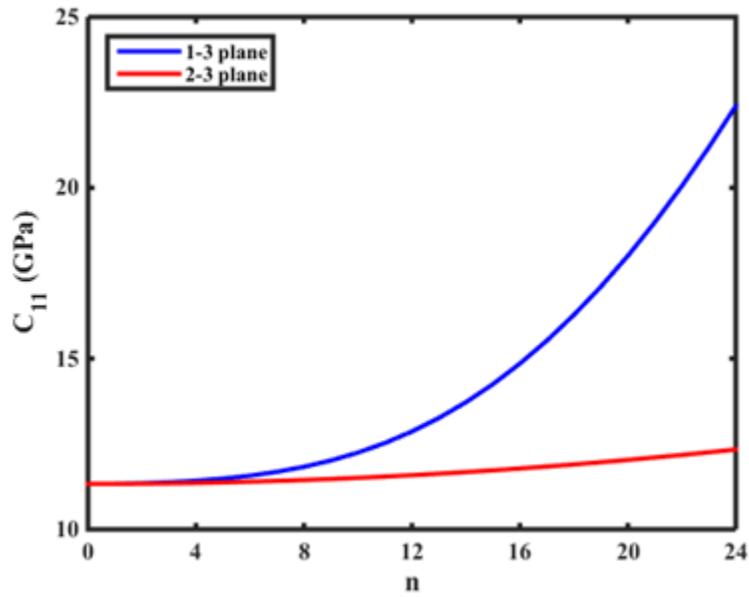

FIG. 27. Effect of waviness of fiber on the effective elastic coefficient $C_{11}$ of the composite ($\omega = n\pi/L_n$; n = 0 to 24).

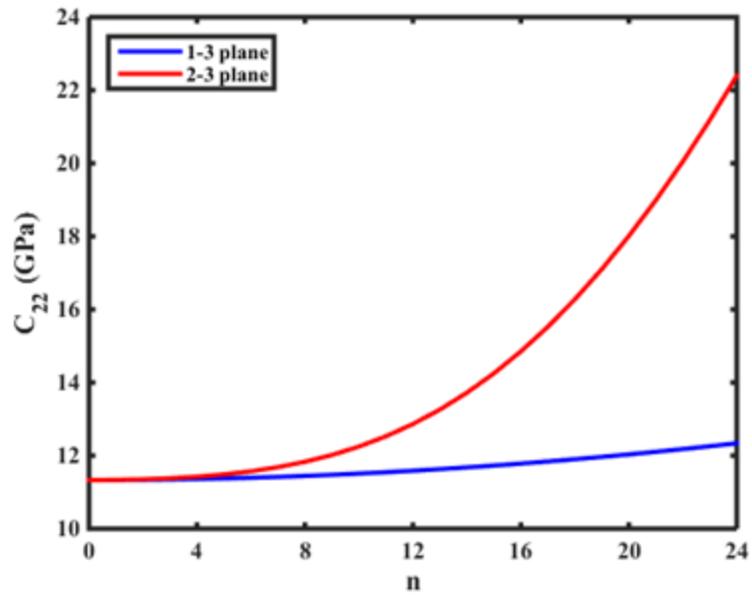

FIG. 28. Effect of waviness of fiber on the effective elastic coefficient $C_{22}$ of the composite ($\omega = n\pi/L_n$; n = 0 to 24).

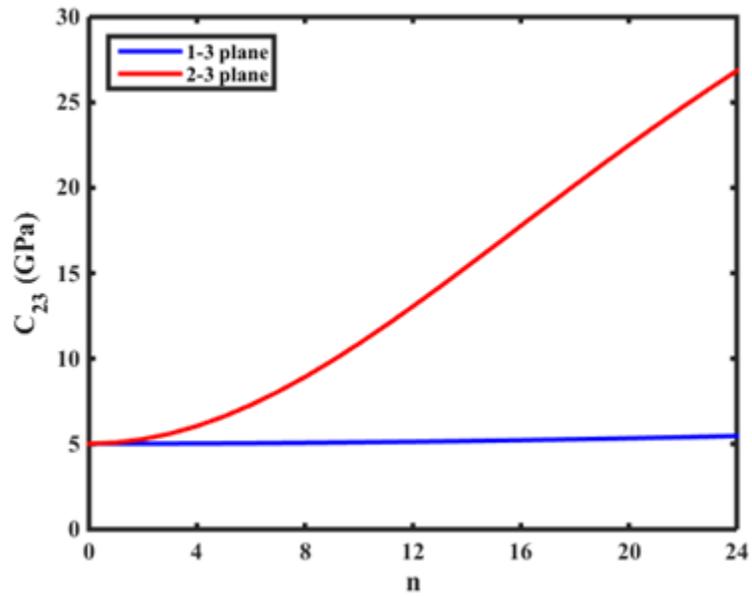

FIG. 29. Effect of waviness of fiber on the effective elastic coefficient $C_{23}$ of the composite ($\omega = n\pi/L_n$; n = 0 to 24).

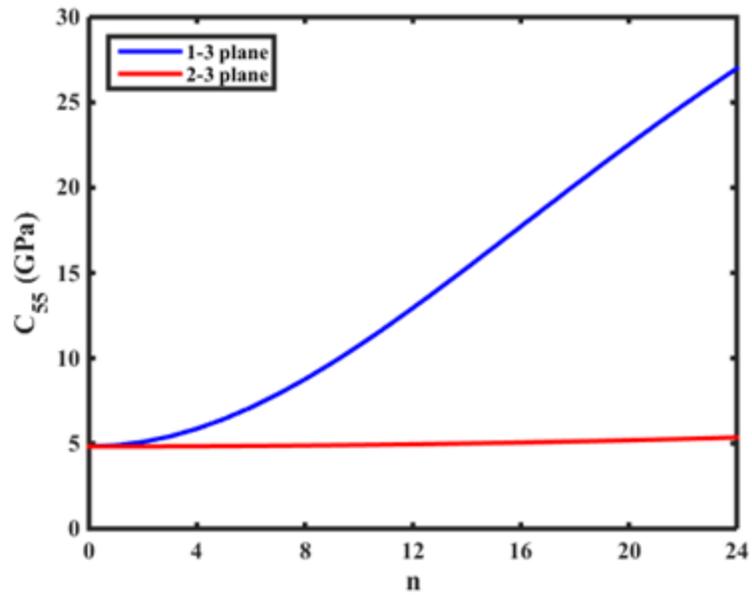

FIG. 30. Effect of waviness of fiber on the effective elastic coefficient $C_{55}$ of the composite ($\omega = n\pi/L_n$; n = 0 to 24).

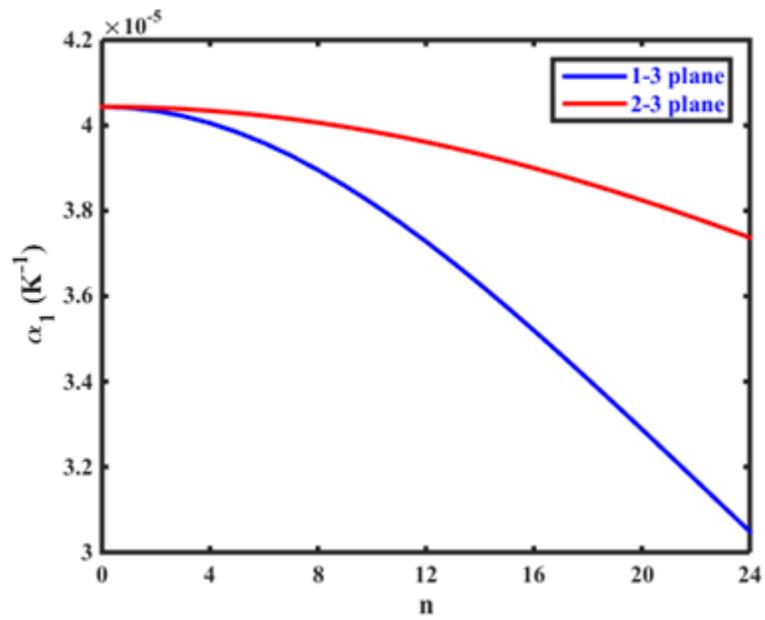

FIG. 31. Effect of waviness of fiber on the effective axial CTE $\alpha_1$ of the composite ($\omega = n\pi/L_n$; n = 0 to 24).

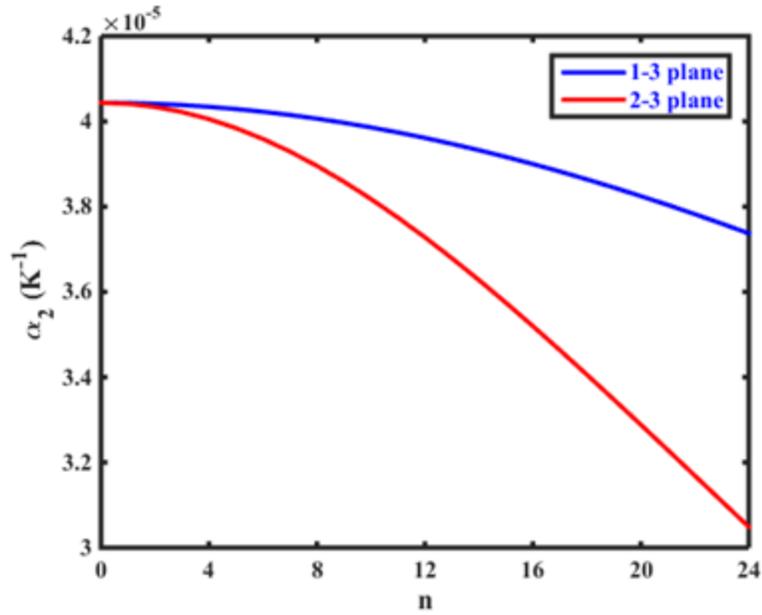

FIG. 32. Effect of waviness of fiber on the effective axial CTE $\alpha_2$ of the composite ($\omega = n\pi/L_n$; n = 0 to 24).

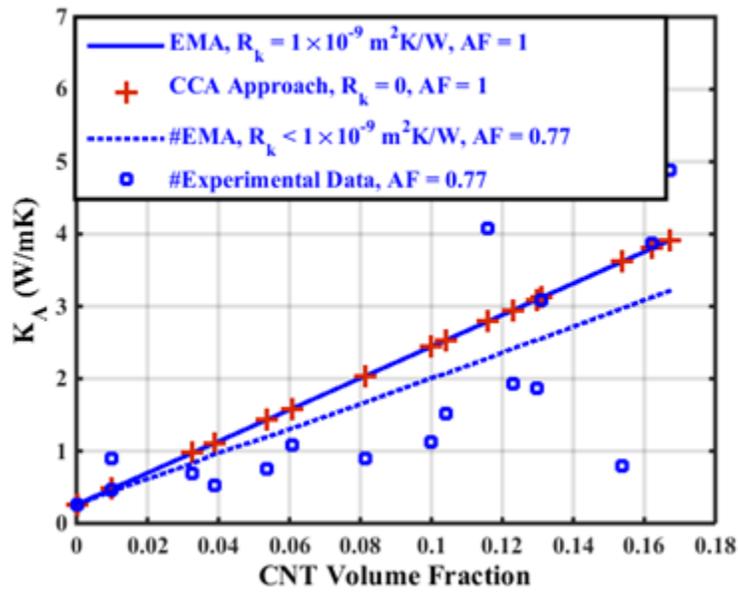

FIG. 33. Comparisons of the variations of the effective axial thermal conductivities of the aligned CNT-reinforced polymer nanocomposite ([#][130]).

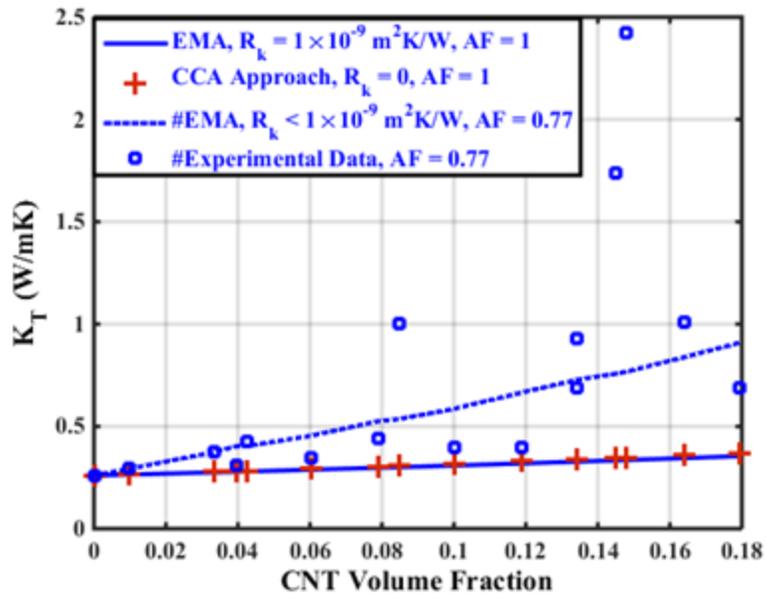

FIG. 34. Comparisons of the variations of the effective transverse thermal conductivities of the aligned CNT-reinforced polymer nanocomposite ([#][130]).

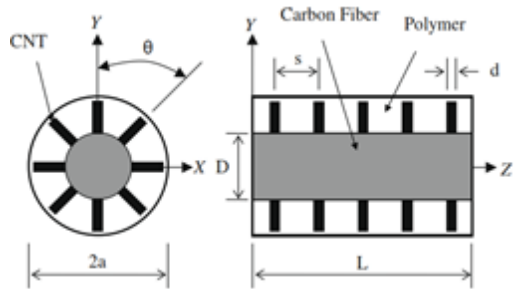

FIG. 35. Transverse and longitudinal cross sections of the NRLC [111].

TABLE 1. Parameters to be used in Halpin-Tsai Eqs. (130) and (131) to predict the effective elastic properties of composite.

| P | $P^f$ | $P^m$ | $\xi$ | Comment |
|---|---|---|---|---|
| $G_{12}$ | $G^f$ | $G^m$ | 1 | Unidirectional short fiber composite |
| $G_{23}$ | $G^f$ | $G^m$ | $\dfrac{1+\nu^m}{3-\nu^m-4(\nu^m)^2}$ | Unidirectional short fiber composite |
| $K_{23}$ | $K^f$ | $K^m$ | $\dfrac{1-\nu^m-2(\nu^m)^2}{1+\nu^m}$ | Unidirectional short fiber composite |
| G | $G^f$ | $G^m$ | $\dfrac{7-5\nu^m}{8-10\nu^m}$ | Particulate composite |
| K | $K^f$ | $K^m$ | $\dfrac{2(1-2\nu^m)}{1+\nu^m}$ | Particulate composite |

TABLE 2. Material properties of the constituent phases of the composite.

| Material | $C_{11}$ (GPa) | $C_{12}$ (GPa) | $C_{44}$ (GPa) | $\alpha_1$ ($10^{-6}$ K$^{-1}$) | (µm) |
|---|---|---|---|---|---|
| Silicon carbide fiber [85, 86] | 487.61 | 153.9 | 166.81 | 2.4 | $d_f = 150$ |
| Fiber-matrix interphase ($\eta = 50$) | 16.05 | 6.36 | 4.84 | 50.52 | t = 2 |
| Epoxy matrix [85] | 6.57 | 3.38 | 1.59 | 51.15 | - |

TABLE 3. Material properties of the fiber-matrix interphases.

| Adhesion exponent | Thickness (μm) | $C_{11}$ (GPa) | $C_{12}$ (GPa) | $C_{44}$ (GPa) | $\alpha_1$ ($10^{-6}$ K$^{-1}$) |
|---|---|---|---|---|---|
| $\eta = 5$ | 1 | 86.76 | 28.5 | 29.13 | 43.14 |
| $\eta = 5$ | 5 | 86.82 | 28.52 | 29.15 | 43.58 |
| $\eta = 10$ | 1 | 50.32 | 17.1 | 16.62 | 46.86 |
| $\eta = 10$ | 5 | 50.4 | 17.12 | 16.63 | 47.41 |

TABLE 4. Comparison of Young's moduli of the CNT-polymer matrix interphase.

| CNT Type | Diameter of CNT (nm) | $h_I$ (nm) | E (Gpa) Molecular Dynamics Simulation [90] | E (GPa) Present Model$^*$ ($\eta = 40$) |
|---|---|---|---|---|
| (10, 0) | 0.78 | 0.3333 | 19.23 | 21.90 |
| (14, 0) | 1.1 | 0.3236 | 17.95 | 17.92 |
| (18, 0) | 1.42 | 0.3158 | 17.90 | 15.65 |

$^*$ In this validation example, Young's modulus ($E^p$) and the Poisson's ratio ($\nu^p$) of the polymer matrix are consedered as 3.8 GPa and 0.4, respectively [110].

TABLE 5. Comparison of the engineering constants of the CNT-reinforced nanocomposite containing sinusoidally wavy CNTs.

| $w = A/\lambda$ | $E_{yy}$ (GPa) | | $E_{xx}$ (GPa) | |
|---|---|---|---|---|
| | FE Model [128] | MT Model | FE Model [128] | MT Model |
| 0 | 18 | 17.800 | 4.577 | 4.448 |
| 0.005 | 17.920 | 17.790 | 4.574 | 4.448 |
| 0.010 | 17.780 | 17.800 | 4.570 | 4.446 |
| 0.015 | 17.690 | 17.750 | 4.566 | 4.445 |
| 0.020 | 17.570 | 17.720 | 4.563 | 4.442 |
| 0.025 | 17.480 | 17.680 | 4.561 | 4.440 |
| 0.030 | 17.130 | 17.630 | 4.559 | 4.435 |

Farsadi et al. [128]: $E^n = 1030$ GPa, $\nu^n = 0.063$, $E^p = 3.8$ GPa, $\nu^p = 0.4$ and CNT volume fraction, $v_n = 0.014$; where w is the waviness ratio; A and $\lambda$ are the amplitude and the wave length of CNT wave; $E^n$ and $\nu^n$ are the Young's modulus and Poisson's ratio of the CNT. $E_{yy}$ and $E_{xx}$ are the axial and transverse Young's moduli of the nanocomposite, respectively.

TABLE 6. Comparison of the CTEs of the CNT-reinforced nanocomposite containing straight CNTs.

| $v_n$ | $\alpha_1 (\times 10^{-6} \text{ K}^{-1})$ | | $\alpha_2 (\times 10^{-6} \text{ K}^{-1})$ | |
|---|---|---|---|---|
| | FE Model [129] | MT Model | FE Model [129] | MT Model |
| 0.5 | 25.2030 | 24.4850 | 69.7540 | 69.9030 |
| 1 | 15.3020 | 15.0720 | 72.8670 | 72.8790 |
| 3 | 5.0978 | 5.1820 | 74.7155 | 74.5120 |
| 5.45 | 2.2670 | 2.2978 | 73.5434 | 73.1220 |
| 7.9 | 1.0643 | 1.1170 | 71.7165 | 71.1300 |
| 10.3 | 0.4253 | 0.4847 | 69.7879 | 69.0140 |
| 15.77 | _0.3201 | _0.2560 | 65.1728 | 64.0370 |

Kirtania and Chakraborty [129]: $E^n = 1000$ GPa, $v^n = 0.2$, $E^p = 3.89$ GPa, $v^p = 0.37$, $\alpha^n = -1.5 \times 10^{-6} \text{K}^{-1}$ and $\alpha^p = 58 \times 10^{-6} \text{K}^{-1}$; where $\alpha_1$ and $\alpha_2$ are the axial and transverse CTEs of the nanocomposite with the straight CNTs, respectively; $\alpha^n$ and $\alpha^p$ are the CTEs of the CNT and the polymer matrix, respectively.

TABLE 7. Comparison of the effective engineering constants of the NRLC [111] with those of the CNT-reinforced nanocomposite.

| | NRLC (2% CNT and 41% IM7 Carbon Fiber) | | MT Model |
|---|---|---|---|
| | Numerical | Experimental | |
| $E_x$(GPa) | 13.93 | 10.02 | 11.91 |
| $v_{xy}$ | 0.34 | – | 0.38 |
| $v_{zx}$ | 0.16 | – | 0.18 |